\documentclass[journal,12pt,draftclsnofoot,onecolumn]{IEEEtran}
\IEEEoverridecommandlockouts
\usepackage{cite}

\usepackage{makeidx,epsfig}
\usepackage{setspace,graphicx,multirow}
\usepackage{amsmath,amssymb,amsfonts}
\usepackage[small]{caption2}

\usepackage{subfigure}
\usepackage{graphicx}
\usepackage{epstopdf}

\usepackage{array}
\usepackage{url}
\usepackage{multirow}
\usepackage{hhline}

\usepackage{xcolor}
\usepackage{textcomp}

\usepackage{bm}
\usepackage{mathtools}
\usepackage{stfloats}
\usepackage{algorithm, algorithmic}
\usepackage{varwidth}

\makeatletter
\def\BState{\State\hskip-\ALG@thistlm}
\makeatother
\usepackage{fixltx2e}

\usepackage{enumitem}

\makeatletter
\newcommand\tinyv{\@setfontsize\tinyv{7pt}{9}}
\makeatother

\usepackage{authblk}

\ifodd 1
\newcommand{\com}[1]{\textbf{\color{blue} (COMMENT: #1)}} 

\else

\newcommand{\com}[1]{}
\fi

\begin{document}
	\bibliographystyle{IEEEtran}
	\bstctlcite{IEEEexample:BSTcontrol}
	
	\title{Energy-Efficient Trajectory Design for UAV-Aided Maritime Data Collection in Wind}
	
	\author{Yifan~Zhang, Jiangbin~Lyu,~\textit{Member,~IEEE},
		and~Liqun~Fu,~\textit{Senior Member,~IEEE}
		
		
		\thanks{Part of this work has been presented in GLOBECOM 2020\cite{OUR}. The authors are with the School of Informatics, and Key Laboratory of Underwater Acoustic Communication and Marine Information Technology (Ministry of Education), Xiamen University, China 361005 (email: zyfan@stu.xmu.edu.cn; \{ljb, liqun\}@xmu.edu.cn). \textit{Corresponding Author: Jiangbin Lyu.}}
	}
	
	\maketitle
	
\begin{abstract}
Unmanned aerial vehicles (UAVs), especially fixed-wing ones that withstand strong winds, have great potential for oceanic exploration and research. This paper studies a UAV-aided maritime data collection system with a fixed-wing UAV dispatched to collect data from marine buoys. We aim to minimize the UAV's energy consumption in completing the task by jointly optimizing the communication time scheduling among the buoys and the UAV's flight trajectory subject to wind effect. The conventional successive convex approximation (SCA) method can provide efficient sub-optimal solutions for collecting small/moderate data volume, whereas the solution heavily relies on trajectory initialization and has not explicitly considered wind effect, while the computational/trajectory complexity both become prohibitive for the task with large data volume.
To this end, we propose a new \emph{cyclical trajectory design} framework with tailored initialization algorithm that can handle arbitrary data volume efficiently, as well as a \emph{hybrid offline-online} (HO$^2$) design that leverages convex stochastic programming (CSP) offline based on wind statistics, and refines the solution by adapting online to real-time wind velocity.
Numerical results show that our optimized trajectory can better adapt to various setups with different target data volume and buoys' topology as well as various wind speed/direction/variance compared with benchmark schemes.
In particular, our proposed HO$^2$ design can effectively adapt to random wind variations with feasible and robust online operation, and proactively exploit the wind for further energy savings in both single-buoy and multi-buoy scenarios.
\end{abstract}
	
\begin{IEEEkeywords}
Maritime data collection, unmanned aerial vehicle, wind effect, cyclical trajectory design, stochastic programming (SP), hybrid offline-online (HO$^2$)
\end{IEEEkeywords}


\section{Introduction}
	
%
%
%
%
%
Marine areas cover almost 71\% of the Earth's surface and provide us with vast amount of resources. New technologies to digitalize/intelligentize oceanic exploration and exploitation are fundamentally changing marine science and information network\cite{Visbeck2018Ocean}. More autonomous operations can increase the efficiency of managing maritime network\cite{Maritimenetworks}, while the coordinated use of various unmanned vehicles helps reducing the risk and mission costs\cite{MaritimeMissions}. Recently, unmanned aerial vehicle (UAV) finds wide applications in wireless communication systems as a mobile base station, relay or data collector\cite{Lyu2016Cyclical,WuQingQingUAVTWC,ZengThroughput,ZengEnergy,YangEnergy,UAVZhang,ZhanChengCollection,ZhanTWCdataCollection} (see also the recent tutorial \cite{ZengAccessing} and the references therein). Moreover, thanks to its high mobility, UAV is also a flexible and cost-effective tool that can be applied in a broad spectrum of marine activities including wireless coverage\cite{maritimecoverage}, surveillance\cite{MaritimeSurveillance}, rescue\cite{MaritimeSearch} and data collection\cite{IOTJ21}.

On the premise of collecting data from marine buoys quickly and in real-time, several telemetry activities\cite{MaritimeData} can be carried out for oceanic monitoring, research and exploitation.
Such maritime data collection can be accomplished by satellites, ships and aircrafts\cite{Roach2009Marine}, whereas satellite communication is typically costly and bandwidth-limited while manned ships/aircrafts incur high manpower/mission cost with potential risk.
In light of the above, it is promising to employ UAVs, especially fixed-wing ones that withstand strong winds above sea surface, as agile data collectors.
UAV, with the advantages of flexible deployment and high mobility, can fly closely to the buoys and exploit the good communication channel to wirelessly and swiftly collect large volume of data.
However, such UAV-aided maritime data collection also faces unique challenges.
First, although fixed-wing UAV typically can be fuel-powered and carry heavier payload than rotary-wing UAV, the limited energy onboard is still one of the critical bottlenecks for long-distance and long-endurance operations at sea. Second, the atmospheric drag caused by prominent and stochastic marine winds cannot be ignored, which affects the UAV's flight and mission as well as its energy consumption. To this end, it is of vital importance to jointly optimize the UAV-to-buoys communications and the UAV's trajectory such that the data collection task can be completed with minimum energy consumption subject to wind effect, which motivates this paper.

Note that for terrestrial scenarios, the energy consumption models of fixed-wing or rotary-wing UAVs are proposed\cite{ZengEnergy}\cite{ZengEnergyMin} assuming zero wind speed, based on which the UAV's trajectory can be jointly optimized with the air-to-ground communications under various setups \cite{Trajectoryopt} including data collection (see, e.g., \cite{YangEnergy,UAVZhang,ZhanChengCollection,ZhanTWCdataCollection}), wireless coverage (see, e.g.,\cite{CoMP,MANET,placement,UAVGBSTWC}), etc.
Typically, the objective is to minimize the flight time duration \cite{ZhanTWCdataCollection}\cite{YongCompletionTime} or a cost function of the propulsion/communication energy consumption\cite{ZengEnergyMin}, while completing the communication task.
The resulted problems are typically non-convex and solved sub-optimally by variants of the successive convex approximation (SCA) technique\cite{ZengThroughput}.
However, these SCA-based solutions heavily rely on the trajectory initialization.
Moreover, for fixed-wing UAVs that must maintain a forward motion to remain aloft, the computational complexity and resulted trajectory complexity both become prohibitive for the task of collecting large volume of data from distributed buoys.
Therefore, new design challenges arise when considering prominent marine wind effect and the practical scenario where each buoy has accumulated a large volume of historical data to be collected.

For maritime scenarios, a UAV-aided ocean monitoring network is proposed in \cite{IOTJ21} where a single UAV hovers above the sink nodes (SNs) on sea surface to relay underwater sensing data to a ground base station. In particular, the UAV position is jointly optimized with the SN-to-UAV communications to minimize the transmission delay.
In \cite{SPAC21}, the deployment and optimization of multi-UAV assisted waterway data collection is investigated, where the problem is simplified into establishing the
Sensor Node - Hover Point (SN-HP) association by using graph
theory, followed by establishing the HP-UAV association.
However, the above works \cite{IOTJ21} and \cite{SPAC21} have not explicitly considered wind effect and contiguous UAV trajectory optimization.
In order to prolong the mission duration, the authors in \cite{maritimedatacollection} investigate resonant beam charging-powered UAV-assisted sensing data collection, and jointly optimize the UAV's trajectory and the power of the charging station to maximize the power transmission efficiency, whereas the wind effect and the UAV's energy consumption are not explicitly considered.
Considering wind effect, the author in \cite{NGMinimum} formulates the problem of finding minimum-energy flight paths of the UAV by utilizing or avoiding the wind, while the authors in \cite{Mission} propose a model to solve a mission planning problem in which UAVs must visit some waypoints in the presence of a wind field to maximize the total reward collected and minimize total flight time. 
However, the data collection scenario is not explicitly considered in \cite{NGMinimum} and \cite{Mission}, and the UAV-to-buoys communication is not jointly optimized. 
Finally, to deal with the random line-of-sight (LoS)/non-LoS channel conditions in city environment, the authors in \cite{HYBIDYOU} propose a hybrid offline-online design of the UAV's trajectory and communication scheduling. On the other hand, since there is no obvious obstacle at sea while there exist prominent and/or stochastic marine winds, it is more important for maritime data collection to investigate the impact of real-time wind on the UAV's flight as well as its communication tasks.

Motivated by the above, in this paper, we investigate a UAV-aided maritime data collection system subject to prominent and/or stochastic marine winds, and aim to minimize the UAV's energy consumption in collecting all data from distributed buoys by jointly optimizing the UAV's trajectory and communication time scheduling. Our main contributions are summarized as follows.

\begin{itemize}
	\item We propose a new \textit{cyclical trajectory design} framework that solves the non-convex problem with arbitrary data volume efficiently subject to wind effect.
	Specifically, the proposed UAV trajectory comprises multiple cyclical laps, each responsible for collecting only a subset of data and hence requiring less mission time in each lap, thereby significantly reducing the computational/trajectory complexity.
	Furthermore, a tailored trajectory initialization algorithm of low complexity is proposed to proactively fit the buoys' topology and the wind.
	
	\item We propose a novel \emph{hybrid offline-online (HO$^2$)} optimization method to tackle the stochastic wind variations, by leveraging the statistical and real-time wind velocity in the offline and online phases, respectively. In the offline phase, the UAV's trajectory, airspeed and communication scheduling are jointly optimized prior to its flight based on given wind statistics. More specifically, the impact of wind speed and direction on the communication task and energy consumption is first revealed for the special case with fixed wind.
	Furthermore, we leverage the \textit{convex stochastic programming (CSP)} technique to obtain a refined and more robust solution for the more challenging case with random wind variations.
	Based on the offline optimized path, for each time slot in the online phase, we further optimize the UAV's airspeed and communication scheduling in adaptation to the instantaneous wind velocity and the amount of data collected accumulatively, which can be solved with low complexity and also complements the offline solution for feasible and flexible operation under real-time wind effect. 
    
	\item Numerical results validate our analysis on the influence of wind speed, direction and variation on the communication task and energy consumption.
	 It is shown that the proposed cyclical trajectory design outperforms the conventional one-flight-only scheme in general, in terms of energy consumption, computational time and trajectory complexity. 
	Moreover, compared with other benchmark trajectories, our optimized trajectory can better adapt to various setups with different target data volume and buoys' topology as well as wind conditions (e.g., with or without wind, fixed or random wind, headwind and/or tailwind, etc.).
	In particular, under random wind conditions, our proposed HO$^2$ design can effectively adapt to wind variations with feasible and robust online operation, and proactively exploit the wind for further energy savings in both single-buoy and multi-buoy scenarios.
\end{itemize}	

The rest of the paper is organized as follows. Section \ref{SectionModel} introduces the system model, based on which we formulate an optimization problem in Section \ref{problemformulation} and present the main idea of the proposed HO$^2$ method for solving it. The special case with fixed wind is introduced in Section \ref{fixed wind}. The offline/online phases are designed in Sections \ref{offlinedesign} and \ref{online}, respectively. Finally, simulation results
and discussions are provided in Section \ref{simulation}, followed by conclusions in Section \ref{conclusions}.

\section{System Model}\label{SectionModel}
\begin{figure}[h]
		\centering
		\includegraphics[width=0.65\linewidth]{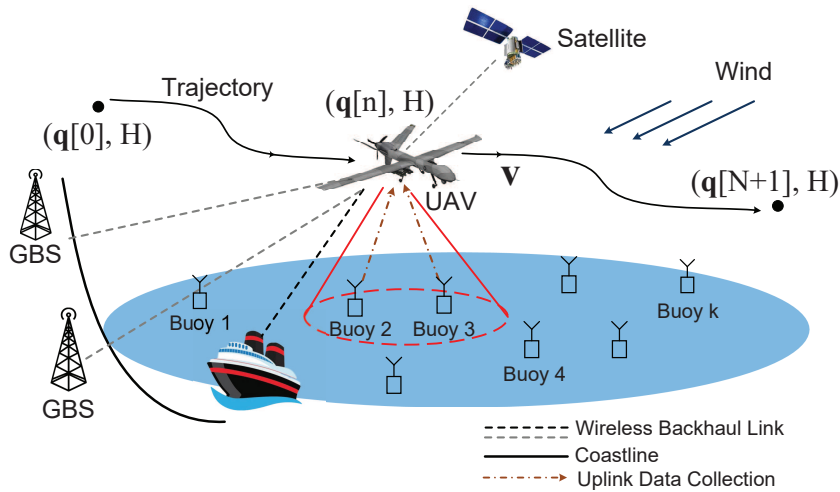}
		\centering
		\caption{Maritime data collection system aided by a fixed-wing UAV in wind.}\label{SystemModel}
\end{figure}

As shown in Fig.\ref{SystemModel}, we consider a maritime data collection system whereby a fixed-wing UAV is dispatched as a mobile data collector to gather information from $K$ buoys.
Denote $\mathbf{b}_k \in {\mathbb {R}^{{\rm{2}} \times {\rm{1}}}}$ as the horizontal location of buoy $k$, $k\in\mathcal{K}\triangleq \{1,\cdots,K\}$, whose total number of information bits to be collected is denoted by ${\bar Q_{{k}}}$. Assume that the UAV has large enough data storage and can store all the collected data locally, which can be downloaded by the pilot after the UAV returns, or transferred back via the backhaul link\footnote{The backhaul link could be provided by the ground base stations (GBS) deployed along the coastline, ships equipped with high-gain antennas or the maritime satellites, as shown in Fig. \ref{SystemModel}.} if real-time data reception is required. Assume that the UAV flies at a constant altitude $H$ meters (m). Denote the UAV's trajectory projected on the horizontal plane by ${\rm{\mathbf{q}}}(t)={\big[x(t_0),y(t_0)\big]^T} \in {\mathbb {R}^{2 \times 1}}$, with $0\le t_0\le T$ and $T$ being the flight time. For simplicity, we discretize the flight time $T$ into $N\!+\!1$ slots, i.e, $n= 1,\cdots,N+1$, each with sufficiently small slot time ${T_s}$. For notation simplicity, we define $\mathcal{N}\triangleq \{0,1,\cdots,N\}$.
As a result, the UAV's trajectory is discretized into $N+2$ waypoints as 
$\mathbf{q}[n]=\mathbf{q}(n{T_s})$, $n\in\mathcal{N}\cup \{N+1\}$. At any waypoint $n$, the distance between the UAV and buoy $k$ is given by $d_k[n]=\sqrt {{H^2} + {{\left\| {\mathbf{q}[n] - {\mathbf{b}_k}} \right\|}^2}} $, $k\in\mathcal{K}$, with $\|\cdot\|$ denoting the Euclidean distance.
	
\subsection{Channel Model}
\indent Since the UAV flies at high altitude and there are no obvious obstacles at sea, we assume for simplicity that the wireless channel between the UAV and buoy $k\in\mathcal{K}$ is dominated by the LoS link which follows the free-space path loss model\footnote{The channel fading can be further considered and a tight lower bound of the expected rate can be derived as in \cite{fadingExpectedrate}, which is a function of the communication distance and hence can be similarly applied in our analytical framework.}, given by
\begin{equation}
	h_k[n] \triangleq {\beta _0}{d_k[n]^{ - 2}} = \frac{{{\beta _0}}}{{{H^2} + {{\left\| {\mathbf{q}[n] - \mathbf{b}{}_k} \right\|}^2}}},
\end{equation}
where ${\beta _{\rm{0}}}$ presents the channel power gain at a reference distance of 1 m. 
Assume that each buoy transmits with power $P_0$. The achievable rate in bits/second (bps) between buoy $k$ and the UAV is given by 
\begin{equation}
	{R_k[n]} \triangleq B{\log _2}\bigg(1 + \frac{{{\gamma _0}}}{{{H^2} + {{\left\| {\mathbf{q}[n] - {\mathbf{b}_k}} \right\|}^2}}}\bigg),\label{rate}
\end{equation}
where $B$ denotes the channel bandwidth in hertz (Hz) and ${\gamma _{\rm{0}}} \triangleq {P_{\rm{0}}}{\beta _{\rm{0}}}{\rm{/}}( {{\sigma ^{\rm{2}}}\Gamma } )$ is defined as the received signal-to-noise ratio (SNR) at a reference distance of 1 m, with ${\sigma ^{\rm{2}}}$ being the noise power at the receiver and $\Gamma$$>$1 as the channel capacity gap caused by practical modulation and coding.
	
\subsection{Multiple Access Scheme}
Assume that the cyclical time-division multiple access (TDMA) protocol\cite{Lyu2016Cyclical} is applied for the UAV to collect data from the $K$ buoys. At any time instant, each buoy is scheduled to communicate with the UAV only when the UAV is close enough. The UAV-enabled cyclical multiple access scheme substantially shortens the communication distance with all buoys by exploiting the mobility of the UAV, thereby improving the system throughput. For each time slot $n$, with cyclical TDMA among the $K$ buoys, denote ${\tau}_{k}[n] \ge 0$ as the allocated time for the UAV to collect data from buoy $k$.\footnote{Besides the data channel, we assume a separate control channel for the UAV to disseminate the time scheduling to the buoys.}
Then we have
\begin{equation}
	\textstyle{\sum}_{k = 1}^K {\tau}_{k}[n]\le T_s^{}, n=1,\cdots,N+1.\label{allocatedtime}
\end{equation}
\indent Therefore, the total amount of information bits collected from buoy $k$ is a function of the UAV's trajectory $\{\mathbf{q}[n]\}$ and allocated time $\{{\tau}_{k}[n]\}$, which is given by\footnote{For notation simplicity, we use $\{x\}$ to denote the set of variables $x$.}
\begin{equation}
	{Q_k}\big( {\left\{ {\mathbf{q}[n]} \right\},\left\{ {{\tau}_{k}[n]} \right\}} \big) = \sum\limits_{n = 1}^{N+1} {{\tau}_{k}[n]{R_k}[n-1]} 
	= B\sum\limits_{n = 1}^{N+1} {{\tau}_{k}[n]} {\log _2}\bigg(1 + \frac{{{\gamma _0}}}{{{H^2} + {{\left\| {{\mathbf{q}[n-1]}-{\mathbf{b}_k}} \right\|}^2}}}\bigg).\label{throughput}
\end{equation}%
Note that since each time slot $n$ spans the waypoints $n-1$ and $n$, for small enough slot time $T_s$, we assume for simplicity that the communication rate at waypoint $n-1$ is adopted for the whole time slot $n$ as in \eqref{throughput}.
As a result, to complete the data collection from buoy $k$, we must have
\begin{equation}
	{Q_k}\big( {\left\{ {\mathbf{q}[n]} \right\},\left\{ {{\tau}_{k}[n]} \right\}} \big)\ge {\bar Q_{{k}}},\forall k.\label{throughputconstraint}
\end{equation}

\subsection{Energy Consumption Model for Fixed-Wing UAV}
	
	
The UAV's energy consumption mainly consists of two parts, namely the communication energy and the propulsion energy. The communication energy includes that for communication circuitry and signal transmission/reception, which is much smaller than the propulsion energy\cite{ZengEnergy} and thus ignored in this paper. For level flight, the instantaneous propulsion energy required by a fixed-wing UAV with air velocity $\mathbf{v}$ and acceleration $\mathbf{a}$ is given by\cite{ZengEnergy}
\begin{equation}
E\big(\{ {\mathbf{v}[n]}\},\{ {\mathbf a[n]}\}\big)\approx \sum\limits_{n = 1}^{N} \left( {{{w_1}{{\left\| {\mathbf{v}}[n]\right\|}^3} + \frac{w_2}{{\left\| {\mathbf{v}[n]} \right\|}}\left( {1 + \frac{{{{\left\| {\mathbf{a}[n]} \right\|}^2}{{}}}}{{{g^2}}}} \right)}} \right)T_s+{\Delta}, \label{energyconsumption}
\end{equation}
where ${\Delta} \triangleq \frac{1}{2}m\left( {{{\left\| {\mathbf{v}[N\!+\!1]} \right\|}^2}-{{\left\| {\mathbf{v}[0]} \right\|}^2}} \right)$, $g$ is the gravitational acceleration with nominal value of 9.8 m/s$^2$, $m$ is the mass of the UAV, and $w_1$ and $w_2$ are two parameters related to the aircraft’s weight, wing area, air density, etc.
	
It can be seen from \eqref{energyconsumption} that the airspeed $\mathbf{v}$ should not be too small, otherwise the required energy to keep the fixed-wing UAV aloft would increase dramatically. In addition, the acceleration $\mathbf{a}$ should not be too large to cause sudden acceleration/deceleration that would consume a lot of energy to generate the required thrust.	
	
\subsection{Wind Effect}\label{windeffect}
\begin{figure}[h]
		\centering
		\includegraphics[height=0.27\linewidth,width=0.78\linewidth]{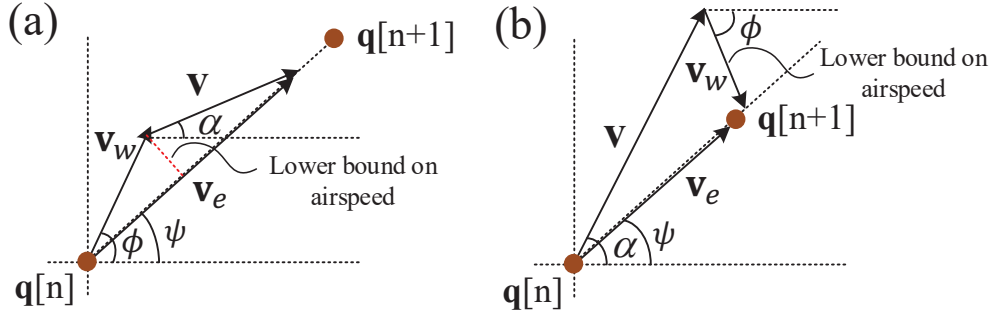}
		\caption{Illustration of (a) tailwind and (b) headwind using the wind velocity ${\mathbf{v}_w[n]}$, as well as the UAV's airspeed ${\mathbf{v}[n]}$ and ground velocity ${\mathbf{v}_e[n]}$.\vspace{-2ex}}
		\label{WindEffect}
\end{figure}

Since $\mathbf{v} \triangleq \dot{\mathbf{q}}$ and $\mathbf{a} \triangleq \dot {\mathbf{v}}$ are respectively the time-varying velocity and acceleration vectors associated with the trajectory point $\mathbf{q}$ in the case of zero wind, for small time step $T_s$, we have the following results based on the first- and second-order Tayler approximations, i.e.,
\begin{equation}\label{v}
\mathbf{v}[n+1] \approx \mathbf{v}[n] + \mathbf{a}[n]{T_s}, n\in\mathcal{N},
\end{equation}
\begin{equation}\label{q}
\mathbf{q}[n+1] \approx \mathbf{q}[n] + \mathbf{v}[n]{T_s} + \frac{1}{2}\mathbf{a}[n]{T_s^2}, n\in\mathcal{N}.
\end{equation}
\indent The effect of wind can be regarded as a shift in the UAV's frame of reference by the wind velocity ${\mathbf{v}_w}\triangleq V_w\angle \phi$, with $V_w$ denoting its absolute value and $\phi$ denoting its angle with the positive horizontal axis, as shown in Fig. \ref{WindEffect}. In general, the wind velocity during the considered period follows a certain random process. For the purpose of exposition, we assume that ${\mathbf{v}_w}$ is a wide-sense stationary process with a certain mean $\mathbb{E}({\mathbf{v}_w})$, standard deviation $\sigma_f$ and correlation parameter $\rho_c$\cite{GPML}.\footnote{Our proposed method is general and can be extended to account for other wind velocity distributions.} The ground velocity $\mathbf{v}_e[ n ]$ at waypoint $n$ can thus be expressed in vector form as
\begin{equation}\label{vwind}
	{\mathbf{v}_e[n]} = {\mathbf{v}[n]} + {\mathbf{v}_w[n]}, n\in\mathcal{N}\cup \{N+1\},
\end{equation}
as shown in Fig. \ref{WindEffect}, where $\alpha$ and ${\psi}$ denote the angles of the UAV's airspeed and ground velocity with regard to the positive horizontal axis, respectively. Accordingly, the acceleration to ground can be expressed as
\begin{equation}\label{aWind}
\mathbf{a}_e[n] \approx \frac{\mathbf{v}_e[n+1]-\mathbf{v}_e[n]}{T_s}, n\in\mathcal{N},
\end{equation}
Since the actual flight path is in the same frame of reference as the ground velocity, equation \eqref{q} is re-written as
\begin{equation}\label{qinWind}
	\mathbf{q}[n+1] \approx \mathbf{q}[n] + \mathbf{v}_e[n]{T_s} + \frac{1}{2}\mathbf{a}_e[n]{T_s^2}, n\in\mathcal{N}
\end{equation}
where the airspeed and acceleration in \eqref{q} is replaced by the ground velocity and acceleration to ground.
If the angle between the wind direction and the flight direction is less than 90$^{\circ}$, i.e., $\left| {\phi-{\psi}} \right|$ $<$90$^{\circ}$, the wind is called \textit{tailwind}.
The tailwind promotes the UAV motion by increasing its ground velocity, so that the UAV can fly a longer distance in a slot time $T_s$. In other words, given fixed flight distance and time, the UAV only needs a lower airspeed to fly through the distance in tailwind. On the contrary, the \textit{headwind} hinders the UAV's forward motion.
	
\begin{figure}
		\centering
		\includegraphics[height=0.25\linewidth,width=0.52\linewidth]{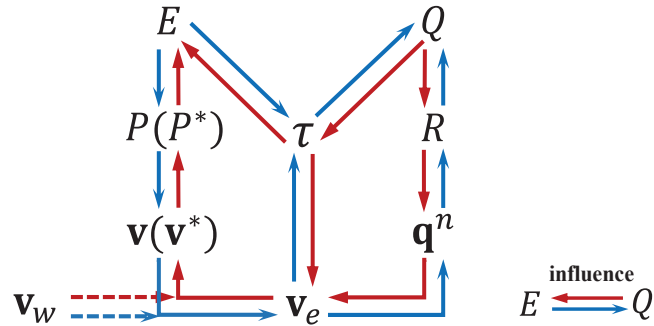}
		\caption{The logical relationship between the variables.\vspace{-2ex}}
		\label{Wind}
\end{figure}
	
\subsubsection{Effect on Communication Task and Energy Consumption}\label{wind effect}
\indent There are two objectives to be optimized in our task, namely the energy consumption $E$ and data volume $Q$, which affect each other through some intermediate variables as shown in Fig. \ref{Wind}. Substituting \eqref{vwind} into \eqref{qinWind}, the wind affects the UAV's ground velocity $\mathbf{v}_e$ and hence its trajectory $\mathbf{q}$, which in turn affects the communication rates $R$ with the buoys given by \eqref{rate} and hence also the communication time scheduling $\tau$ among the buoys, thereby affecting the overall energy consumption given by \eqref{energyconsumption} in completing the data collection task given by \eqref{throughputconstraint}. 
More specifically, when the UAV gets close to the buoy, the headwind reduces the ground velocity and shortens the flight distance in each time slot, allowing the UAV to maintain a good communication channel to collect data. In addition, after finishing the data collection task, the tailwind increases the ground velocity and thereby helps the UAV to reach the next trajectory point quickly. 
On the contrary, adversarial effect can happen if the UAV encounters wind of the reverse direction in the above scenarios.

To see the wind effect on the energy consumption, we first consider the special case where the UAV has a constant velocity with zero acceleration in a short time slot. In this case, the energy function (\ref{energyconsumption}) is a convex function and there exists an optimal velocity ${\mathbf{v}}^{\rm{*}}$ in achieving the minimum propulsion power ${P}^{\rm{*}}$, while too fast or too slow airspeed will lead to increased energy consumption. 
For a given communication task that requires a certain profile of $\{\mathbf{v}_e\}$ and $\{\tau\}$, the wind velocity $\mathbf{v}_w$ affects the airspeed ${\mathbf{v}}={\mathbf{v}_e}-{\mathbf{v}_w}$ and thus affects the UAV's energy consumption.
Therefore, it is of great importance to jointly optimize the UAV's trajectory, airspeed and communication scheduling such that the wind effect can be properly utilized for the UAV to complete its mission with high energy efficiency.
	
\subsubsection{Constraints on Minimum Airspeed}
There are two constraints on the UAV’s minimum airspeed. First, there is a minimum airspeed for the UAV to maintain level flight, which is known as the \textit{stall speed} and denoted by $V_s$. 
Second, the UAV needs to fly from $\mathbf{q}[n]$ to $\mathbf{q}[n+1]$ within a slot time $T_s$ subject to the wind effect.
In the case with tailwind as shown in Fig. \ref{WindEffect}(a), it is required that $||\mathbf{v}[n]|| \ge{V_w}\left| {\sin (\phi  - {\psi})} \right|$.
On the other hand, in the case with headwind as shown in Fig. \ref{WindEffect}(b), it is required that $||\mathbf{v}[n]||\ge{V_w}$.

\section{Problem Formulation and Hybrid Offline-Online Design}\label{problemformulation}
Based on the above system model, for the considered UAV-aided maritime data collection problem, we aim to minimize the UAV's energy consumption in collecting the required data volume $\bar Q_k$ from each of the $K$ buoys, by jointly optimizing the communication time scheduling among the buoys and the UAV's flight trajectory subject to wind effect. The problem can be formulated as problem (P1), i.e.,
\begin{align}
\mathrm{(P1)}:
\underset{
\begin{subarray}{c}
\{ {\mathbf{q}[n]}\} ,\{ {\mathbf{v}[n]}\},\\
\{ {\mathbf{a}[ n ]}\},\{ {\tau_{k}[n]}\}\notag\\
\end{subarray}
}{\min}& E\big(\{ {\mathbf{v}[n]}\},\{ {\mathbf a[n]}\}\big)\\	
\text{s.t.}\quad 
    &\eqref{allocatedtime}, \eqref{throughputconstraint}, \eqref{v}, \eqref{qinWind},\notag\\
	&{\mathbf{q}[0]} = {\mathbf{q}_0}, {\mathbf{q}[N+1]} = {\mathbf{q}_F},\label{q0qF}\\ 
	&{\mathbf{v}[0]} = {\mathbf{v}_0}, {\mathbf{v}[N+1]} = {\mathbf{v}_F},\label{v0vF}\\ 
	&\left\| {\mathbf{a}[n]} \right\| \le {a_{\max }}, n\in\mathcal{N},\label{aconstraint}\\
	&{\tau_{k}[n]} \ge 0,\forall k, n=1,\cdots,N+1,\label{timeslot}\\
	&{V_{\min}} \le \left\| {{\mathbf{v}}[n]} \right\| \le {V_{\max }}, n\in\mathcal{N}\setminus\{0\},\label{vconstraint}
\end{align}
where $E\big(\{ {\mathbf{v}[n]}\},\{ {\mathbf a[n]}\}\big)$ denotes the energy consumption in \eqref{energyconsumption}, ${\mathbf{q}_{0}}$ and ${\mathbf{q}_{F}} \in {\mathbb {R}^{{\rm{2}} \times {\rm{1}}}}$ represent the UAV's initial and final locations projected onto the horizontal plane, respectively; ${V_{\max }}$ and ${a_{\max }}$ represent the maximum speed and acceleration, respectively; and ${V_{\min}}$ represents the minimum airspeed subject to the two constraints discussed in \ref{windeffect}. For simplicity, we choose ${V_{\min}}$ based on the upper bound of these two constraints, i.e., ${V_{\min}}=\max$$\{V_w, V_s\}$.

\begin{figure}[h]
		\centering
		\includegraphics[height=0.44\linewidth,width=0.76\linewidth]{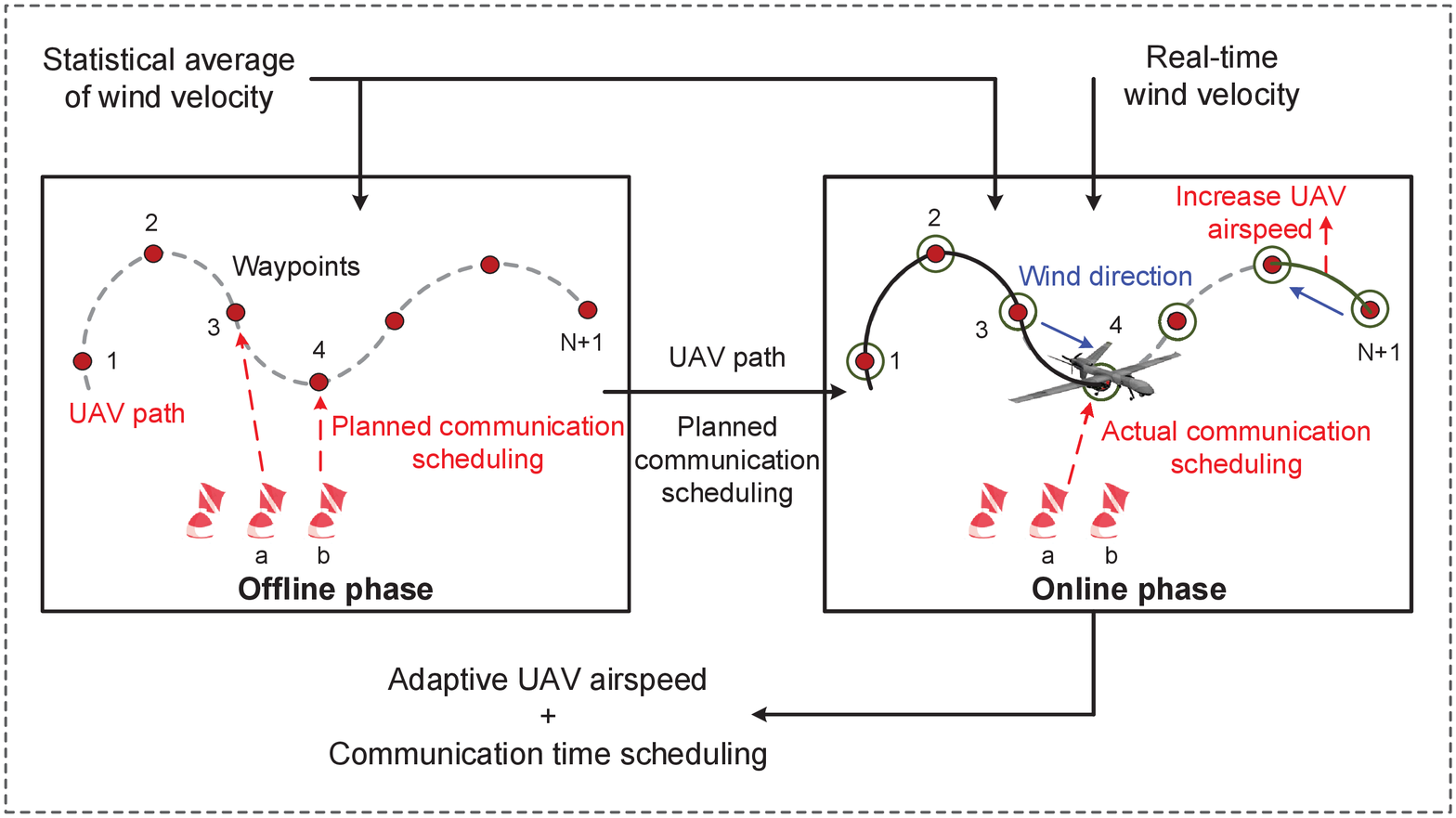}
		\caption{The proposed HO$^2$ optimization for UAV-aided maritime data collection in wind.\vspace{-2ex}}
		\label{hybrid}
\end{figure}

The optimal solution to problem (P1), in general, is difficult to obtain due to the lack of the exact wind velocity information at each time slot prior to flight. To address this difficulty, we assume that the wind statistics are known for the considered time period, and the instantaneous wind velocity can be obtained in real time along the UAV's flight. Thereby, we propose a novel and general method to obtain a suboptimal solution to problem (P1), called \emph{hybrid offline-online} (HO$^2$) optimization, by leveraging both the statistical and real-time wind velocity information. Our proposed method consists of the following two phases as illustrated in Fig. \ref{hybrid}, which are briefly described as follows and will be elaborated in more details in the subsequent sections.
	
\subsubsection{Offline phase}
Prior to the UAV’s flight, we design an initial UAV trajectory and communication scheduling policy based on the statistical distribution of wind velocity. The policy is computed offline by solving an optimization problem to minimize the UAV's energy consumption in collecting the required data volume from marine buoys. The resultant UAV trajectory yields a statistically favorable UAV path that is specified by a sequence of ordered waypoints and line segments connecting them.
	
\subsubsection{Online phase}
In the online phase, we set the UAV trajectory (waypoints) as that obtained from the offline phase. Then, for the online flight from ${\mathbf{q}[n]}$ to ${\mathbf{q}[n+1]}$, we formulate a convex problem to minimize the UAV's current energy consumption subject to the scheduled data collection volume in the current line segment, by slightly adjusting the UAV's trajectory and optimizing airspeed and communication scheduling with buoys in adaptation to the current wind velocity.
	
Note that different from prior works on UAV trajectory and communication design which neglect the wind effect, especially the variance of wind in real time, our proposed method decouples the design into the UAV trajectory optimization in the offline phase, followed by the real-time adjustment of the UAV flying speed and communication time scheduling in the online phase. This is motivated by the fact that the path optimization to complete the whole task requires solving a complex non-linear optimization problem (as will be detailed in Sections \ref{fixed wind} and \ref{offlinedesign}) which is desired to be implemented offline; while the online adaptation can be optimized by solving a simpler convex problem for the current time slot to complete the flight between two planned waypoints under real-time wind condition, which requires low computational complexity and thus can be implemented by the UAV in real time. 

Our proposed hybrid design is expected to achieve good performance owing to the following reasons. First, the offline phase ensures a statistically favorable UAV trajectory that well considers the statistical  wind effect for minimizing the UAV's energy consumption. Second, our proposed online design, built upon the trajectory obtained offline, endows the UAV with self-adaptation to the actual wind velocity to further improve the performance. As discussed in \ref{wind effect} about the wind effect on energy consumption and communication tasks, it is beneficial to optimize the UAV's airspeed and the communication time scheduling at each line segment according to real-time wind conditions. For instance, as illustrated in Fig. \ref{hybrid}, suppose that the UAV is scheduled to collect data from buoys $a$ and $b$ at waypoints $3$ and $4$, respectively. If the UAV encounters additional wind at waypoint $3$ in real time which causes the UAV to deviate from its original trajectory and reach a position farther away from buoy $a$, then it still needs to communicate with buoy $a$ at waypoint $4$ to collect enough data. Besides adjusting the communication scheduling, it can also adjust its airspeed to reach a suitable position to collect the targeted data volume planned offline. For example, if the UAV is far away from the buoys and encounters headwind, it can increase airspeed to reach next waypoint quickly. 

In the following sections, we first introduce the UAV trajectory design under fixed wind and also propose an offline design based on given distribution of random wind. By leveraging the above designs as initial scheduling/trajectory, we then introduce the online phase to adjust the UAV's airspeed and communication time scheduling according to the real-time wind velocity.

\section{Special case with fixed wind} \label{fixed wind}
To better analyze the effect of wind on the UAV's trajectory/resulted energy consumption and investigate how to utilize the wind to complete the task with lower energy consumtion, we first consider the fixed wind situation, i.e., $\mathbf{v}_w$ is constant for all time slots. The obtained solution can also provide a good initial trajectory for the hybrid design. 
	
Even when $\mathbf{v}_w$ is fixed, problem (P1) still requires joint optimization of the trajectory $\{ {\mathbf{q}[n]} \}$, airspeed $\{\mathbf{v}[n]\}$, acceleration $\{\mathbf{a}[n]\}$ and communication scheduling $\{ {{\tau_{k}[n]}} \}$, which are coupled with each other through the cost function as well as constraints \eqref{allocatedtime}, \eqref{throughputconstraint}, \eqref{v} and \eqref{qinWind}. Due to the non-convex constraints \eqref{throughputconstraint} and \eqref{vconstraint}, and the non-convex cost function $E\big(\{ {\mathbf{v}[n]}\}, \{ {\mathbf{a}[n]}\} \big)$ in \eqref{energyconsumption} for the fixed-wing UAV's energy consumption, problem (P1) is complicated and cannot be directly solved by using standard convex optimization techniques. 
To tackle this problem, we leverage the SCA technique as the basic optimization tool, which approximates each non-convex function involved in (P1) by a convex and differentiable function based on the first-order Taylor approximation at a certain local point\cite{ZengEnergy}. A sub-optimal solution to the original non-convex problem can then be obtained by solving a series of convex sub-problems with successively updated local points at each iteration. 
Assuming that the initial and final airspeed of the UAV are equal, we have ${\Delta}=0$. 
In addition, to deal with the non-convex constraint \eqref{vconstraint} and the objective function, we reformulate (P1) by introducing slack variables $\{ {\delta _n}\}$ as
\begin{align} 
	\mathrm{(P2.1)}:& \underset{
		\begin{subarray}{c}
		\{ {\mathbf{q}[n]}\} ,\{ {\mathbf{v}[n]}\},\\
		\{ {\tau_{k}[n]}\},\{ {\mathbf{a}[n]}\},\{ {\delta _n}\}\notag\\
		\end{subarray}
	}{\min} \sum\limits_{n = 1}^{N}  \left( {{{w_1}{{\left\| {\mathbf{v}}[n]\right\|}^3} + \frac{w_2}{{{\delta _n}}}\left( {1 + \frac{{{{\left\| {\mathbf{a}[n]} \right\|}^2}{{}}}}{{{g^2}}}} \right)}} \right)T_s\\
	\text{s.t.}\quad 
	&\eqref{allocatedtime}, \eqref{throughputconstraint}, \eqref{v}, \eqref{qinWind}-\eqref{timeslot},\notag\\
	&{\delta _n} \ge {V_{\min}}, n\in\mathcal{N}\setminus\{0\},\label{delta}\\
	&{\left\| {{\mathbf{v}}[n]} \right\|}^2 \ge \delta _n^2, n\in\mathcal{N}\setminus\{0\},\label{vcons}\\
	&\left\| {{\mathbf{v}}[n]} \right\| \le {V_{\max }}, n\in\mathcal{N}\setminus\{0\}.\label{vmax}
\end{align}

Note that at the optimal solution to (P2.1), we must have ${\delta _n}=\left\| {{\mathbf{v}}[n]} \right\|$, $\forall n$, since otherwise we can always increase ${\delta _n}$ to obtain a strictly larger objective value. Thus, (P2.1) is equivalent to (P1). With such a reformulation, the objective function is now jointly convex with respect to $\{{\mathbf{v}[n],\mathbf{a}[n],\delta _n}\}$, but with the new non-convex constraint \eqref{vcons}. To tackle this difficulty,
notice that the left-hand side (LHS) of \eqref{vcons} is a convex and differentiable function which can be lower bounded by its first-order Taylor approximation at a certain local point $\mathbf{v}^{(l)}[n]$ in the $l$-th iteration, i.e.,
\begin{equation}
		{\left\| {{\mathbf{v}}[n]} \right\|^2} \ge {\left\| \mathbf{v}^{(l)}[n] \right\|^2} + 2\mathbf{v}^{(l)T}[n](\mathbf{v}[n] - \mathbf{v}^{(l)}[n]), n\in\mathcal{N}\setminus\{0\}, \label{vappro}
\end{equation}
where the equality holds at the point ${\mathbf{v}}[n] = \mathbf{v}^{(l)}[n]$. 
Furthermore, at the local point $\mathbf{v}^{(l)}[n]$, both the function ${\left\| {{\mathbf{v}}[n]} \right\|^2}$ and its lower bound have identical gradient. Define the new constraint,
\begin{equation}
	{\left\| \mathbf{v}^{(l)}[n]\right\|^2} + 2\mathbf{v}^{(l)T}[n](\mathbf{v}[n] - \mathbf{v}^{(l)}[n]) \ge \delta _n^2, n\in\mathcal{N}\setminus\{0\}, \label{vapproximation}
\end{equation}
which is convex since its LHS is linear with ${\mathbf{v}}[n]$. Then the inequality \eqref{vappro} shows that the convex constraint \eqref{vapproximation} always implies the non-convex constraint \eqref{vcons}, but the reverse is not true in general.

Similarly, to deal with the non-convex constraint \eqref{throughputconstraint}, we introduce slack variables $\{ {A_k[n]}\}$ with
\begin{equation}
	A_k^2[n] = {\tau _k}[n]{\log _2}\left( {1 + \frac{{{\gamma _0}}}{{{H^2} + {{\left\| {\mathbf{q}[n-1] - {\mathbf{b}_k}} \right\|}^2}}}} \right).
\end{equation}
Therefore, based on similar transforms as in \eqref{vappro} and \eqref{vapproximation}, it suffices to replace \eqref{throughputconstraint} by the following two convex constraints, i.e.,
\begin{equation}\label{throughputconstraint2}
	B\sum\limits_{n = 1}^{N+1} {\left( {{A_k^{(l)2}[n]} + 2{A_k^{(l)}[n]}\left( {{A_k[n]} - {A_k^{(l)}[n]}} \right)} \right)}  \ge {\bar Q_k}, \forall k,
\end{equation}
\begin{equation}\label{Ank}
\frac{{{A_k^2}[n]}}{{{\tau _k}[n]}} \le {R_k^{(l)}[n]}, \forall k, n=1,\dots,N+1,
\end{equation}
where ${A_k^{(l)}[n]}$ is the value of ${A_k[n]}$ at the $l$-th iteration, ${R_k^{(l)}[n]} = {\log _2}\left( {1 + \frac{{{\gamma _0}}}{{{H^2} + {{\left\| {\mathbf{q}{{[n]}^{(l)}} - {\mathbf{b}_k}} \right\|}^2}}}} \right) - {\beta_k[n]}\left( {{{\left\| {\mathbf{q}[n] - {\mathbf{b}_k}} \right\|}^2} - {{\left\| {\mathbf{q}^{(l)}{{[n]}} - {\mathbf{b}_k}} \right\|}^2}} \right)$, with ${\beta _k[n]} = \frac{{({{\log }_2}e){\gamma _0}}}{{\left( {{H^2} + {{\left\| {\mathbf{q}^{(l)}{{[n]}} - {\mathbf{b}_k}} \right\|}^2}} \right)\left( {{{\left\| {\mathbf{q}^{(l)}{{[n]}} - {\mathbf{b}_k}} \right\|}^2} + {\gamma _0}} \right)}}$. 
As a result, problem (P2.1) can be reduced to the following problem, i.e.,
\begin{align}
	\mathrm{(P2.2)}:& \underset{
		\begin{subarray}{c}
		\{ {\mathbf{q}[n]}\} ,\{ {\mathbf{v}[n]}\},\{ {\mathbf{a}[n]}\},\\
		\{ {\tau_{k}[n]}\},\{ {A_{k}[n]}\},\{ {\delta _n}\}\notag\\
		\end{subarray}
	}{\min} \sum\limits_{n = 1}^{N}  \left( {{{w_1}{{\left\| {\mathbf{v}}[n]\right\|}^3} + \frac{w_2}{{{\delta _n}}}\left( {1 + \frac{{{{\left\| {\mathbf{a}[n]} \right\|}^2}{{}}}}{{{g^2}}}} \right)}} \right)T_s\\ 
	\text{s.t.}\quad 
	&\eqref{allocatedtime}, \eqref{v}, \eqref{qinWind}-\eqref{timeslot}, \eqref{delta}, \eqref{vmax}, \eqref{vapproximation}, \eqref{throughputconstraint2}, \eqref{Ank},\notag
\end{align}
which is now a convex problem whose optimal value provides an upper bound for the original problem (P2.1), and its optimal solution can be used as the local point in the next iteration $l+1$.
By successively updating the local point at each iteration via solving (P2.2), we thus obtain an iteratively-refined sub-optimal solution for the non-convex problem (P2.1).

The computational complexity of each convex sub-problem (P2.2) can be shown to grow in the order of $O({(KN)^{3.5}})$ based on similar analysis as in \cite{UAVZhang}, where $N$ is the number of time slots required to complete the data collection task in our considered setup. 
Such a SCA-based solution can be obtained efficiently for small/moderate data volume to be collected, which corresponds to short mission time required and hence small number of time slots $N$ in completing the task.
However, in our considered maritime data collection scenario, it is likely that each buoy stores a large volume of historical maritime/undersea monitoring data, which requires prolonged mission time for the UAV to complete the data collection.
As a result, for a fixed-wing UAV that must maintain a forward motion to remain aloft, the computational complexity and resulted trajectory complexity both become prohibitive for the task of collecting large volume of data from distributed buoys.
In addition, the SCA-based solution heavily relies on the trajectory initialization and may trap in some locally optimal solution.
Moreover, the effect of wind on the data collection task has not been explicitly considered in the literature, which in fact could be exploited to design tailored trajectories for minimizing the UAV's energy consumption in completing the data collection task.
	
To address the above challenges, we propose a new cyclical trajectory design framework that can handle arbitrary data volume efficiently subject to the prominent marine wind effect.
Specifically, the proposed UAV trajectory comprises $M\geq 1$ cyclical laps, each responsible for collecting only a fraction $1/M$ of data and hence requiring less mission time in each lap, thereby significantly reducing the computational/trajectory complexity.
Furthermore, the saved computational effort allows us to design tailored algorithms to search for better trajectory initialization that fits the buoys' topology and the wind.
For the purpose of exposition, we consider two simple patterns of initial trajectories, namely the circular trajectory and the 8-shape trajectory, and assume for simplicity that each buoy has the same data volume $Q$ to be collected.
The detailed algorithms are summarized in Algorithms 1 and 2.

\begin{algorithm}\caption{Cyclical Trajectory Optimization}\label{CyclicalTrajectory}
	\begin{small}
	\textbf{Input:} Buoy locations $\mathbf{b}_k, k\in\mathcal{K}$, wind velocity ${\mathbf{v}_w}$, data volume $Q$ and the number of laps $M$.\\
	\textbf{Output:} Cyclical trajectory $\{{\mathbf{q}^{\rm{*}}[n]}\}$, airspeed $\{{\mathbf{v}^{\rm{*}}[n]}\}$, acceleration $\{{\mathbf{a}^{\rm{*}}[n]}\}$ and communication scheduling $\left\{ {{\tau_{k}^*[n]}} \right\}$.
		
    \begin{algorithmic}[1]
		\STATE Set $Q_0$ = ${Q/M}$ and initialize the maximum number of iterations $l_0$.
		\STATE Choose one pattern of initial trajectory (e.g., circular or 8-shape).
		\STATE Call \textbf{InitialTrajectory}$\big(Q_0, \{\mathbf{b}_k\}, \mathbf{v}_w\big)$ and obtain the time period $T_0$ and initialization $\{\mathbf{q}^0[n]\}$, $\{\mathbf{v}^0[n]\}$ and $\left\{ {{\tau_{k}^0[n]}} \right\}$.\label{StepInitial}
		\STATE Solve problem (P2.1) based on SCA and obtain the solution $\{{\mathbf{q}[n]}\}$, $\{{\mathbf{v}[n]}\}$, $\{{\mathbf{a}[n]}\}$ and $\left\{ {{\tau_{k}[n]}} \right\}$. Record the energy consumption $E$.\label{StepSCA}
		\REPEAT\label{Alg1Repeat}
		\STATE Fine-tune the time period $T$ around $T_0$ (For 8-shape trajectory: search for optimal orientation $\theta$), with corresponding modification on the initialization obtained in Step \ref{StepInitial}.
		\STATE Call step \ref{StepSCA}.
		\UNTIL The optimal solution is found or $l_0$ is reached. 
		\STATE Output the solution with the minimum $E$ recorded.\label{Alg1output}
    \end{algorithmic}
	\end{small}
\end{algorithm}

Note that for large data volume, the communication time that the UAV enters/leaves the cyclical trajectory can be practically ignored.
Our proposed cyclical trajectory optimization in Algorithm \ref{CyclicalTrajectory} consists of two main phases, i.e., the initialization phase (steps 1$\sim$\ref{StepSCA}) and the fine-tuning phase (steps \ref{Alg1Repeat}$\sim$\ref{Alg1output}).
The initialization phase efficiently finds a simple feasible solution of problem (P2.1) by fixing the trajectory pattern (e.g., circular trajectory with one circle of radius $r$ centered at the origin, or 8-shape trajectory with two circles of radius $r$ tangent to each other).
Given the trajectory pattern, the \textbf{InitialTrajectory} procedure searches for the optimal time period $T_0$ and trajectory parameters (e.g., the radius $r$ and/or the orientation $\theta$ of the 8-shape trajectory, illustrated later in Fig. \ref{8-shape}(a)) that achieve the least possible energy consumption within a certain maximum number of iterations (denoted by $l_1$, $l_2$, etc.) while satisfying the throughput constraint.
Note that the throughput feasibility test under a given trajectory pattern only involves simple linear inequalities with the time allocation $\left\{ {{\tau_{k}[n]}} \right\}$, which can be done much faster than solving the original problem (P2.1) using SCA.
Based on the obtained feasible time period $T_0$ and initialization $\{\mathbf{q}^0[n]\}$, $\{\mathbf{v}^0[n]\}$ and $\left\{ {{\tau_{k}^0[n]}} \right\}$, we can then fine-tune the time period $T$ around $T_0$ (for a maximum of $l_0$ iterations), and apply SCA in each iteration to fine-tune the trajectory and time allocation, hence further reducing the energy consumption but with much fewer calls of the SCA routine.

Finally, note that by partitioning into $M$ laps, 
the mission time and hence the number of time slots in each lap can be roughly reduced to $1/M$ of the one-flight-only scheme, whereby the computational complexity can be reduced to around ${1/O({M^{3.5}})}$ of the one-flight-only scheme when SCA is applied.
Furthermore, the saved computational effort allows us to search for better trajectory initialization that fits the buoys' topology and the wind, which is reduced to feasibility tests on linear inequalities and hence involves much lower complexity than SCA.
Moreover, thanks to the reduced time period $T$ and the simpler cyclical trajectory, it is typically much easier to search and fine-tune the mission time and trajectory initialization\footnote{Note that these steps are also needed in the one-flight-only scheme.} before feeding to the SCA routine.
Therefore, our proposed cyclical trajectory design framework can efficiently reduce the UAV's energy consumption in completing the data collection task, especially for large data volume to be collected.
	
\begin{algorithm}\caption{\textbf{InitialTrajectory} Procedure}\label{InitialTrajectory}
    \begin{small}
	$\big[T_0, \{\mathbf{q}^0[n]\}, \{\mathbf{v}^0[n]\}, \left\{ {{\tau_{k}^0[n]}} \right\} \big]$=\textbf{InitialTrajectory}$\big(Q_0, \{\mathbf{b}_k\}, \mathbf{v}_w\big)$
		
	\begin{algorithmic}[1]
		\STATE Assume constant ground speed $V$ (0 $\le$ $V$ $\le$ $V_\textrm{max}$). Set the geometric center of $\{\mathbf{b}_k\}$ as the origin. Initialize $l_1$, $l_2$ and $l_3$.
		\REPEAT 
		\STATE Search for $T_0$.\label{SearchT0}
		\REPEAT 
		\STATE Search for $r$. Obtain $V$ based on 2$\pi r=VT$ for circular trajectory and 2$\pi r$ = ${VT/2}$ for 8-shape trajectory.\label{Searchr}
		\STATE Given $T_0$, $r$, $V$ and $\mathbf{v}_w$, obtain the energy consumption. Find feasible $\left\{ {{\tau_{k}^0[n]}} \right\}$ by solving (P2.1) via linear programming. 
		\STATE If it is infeasible, return to step \ref{Searchr}.
		\UNTIL The optimal solution is found or $l_2$ is reached.
		\STATE If it is infeasible, return to step \ref{SearchT0}.
		\UNTIL $T_0$ that minimizes the energy is found or $l_1$ is reached.
		\STATE Given $T_0$, $r$, $V$ and $\mathbf{v}_w$, obtain $\{\mathbf{q}^0[n]\}$, $\{\mathbf{v}^0[n]\}$ and $\left\{ {{\tau_{k}^0[n]}} \right\}$.	
	\end{algorithmic}
	\end{small}
\end{algorithm}

\section{Offline Design Based on SP}\label{offlinedesign}
The above section describes the UAV's trajectory design under fixed wind. However, in practice, the wind velocity changes randomly which affects the UAV's ground velocity and hence its trajectory and further the communication rate and energy consumption. Therefore, this section aims to offline design the UAV trajectory and communication time scheduling by explicitly considering the random wind variations, such that when the offline solution is applied in real time, the online operation is more robust to wind changes in the sense that the flight and communication constraints are less likely to be violated. To this end, stochastic programming (SP) is leveraged to model the robust optimization problem with uncertain data, i.e., the random wind in our case.

Specifically, by first substituting \eqref{v}, \eqref{vwind} and \eqref{aWind} into \eqref{energyconsumption} and other constraints, problem (P2.2) can be rewritten as
\begin{small}
\begin{align}
	\mathrm{(P3.1)}:& \underset{
		\begin{subarray}{c}
			\{ {\mathbf{q}[n]}\} ,\{ {\mathbf{v}_e[n]}\},\{ {\tau_{k}[n]}\},\\
			\{ {A_{k}[n]}\},\{ {\delta _n}\}\notag\\
		\end{subarray}
	}{\min}{\sum\limits_{n=1}^{N}{\left({{w_1}{{\left\| {{\mathbf{v}}[n]} \right\|}^3}+\frac{{{w_2}}}{{{\delta _n}}}+\frac{{{w_2}{{\left\| {{\mathbf{v}}[n+1]-{\mathbf{v}}[n]} \right\|}^2}}}{{{{\left( {g{T_s}} \right)}^2}{\delta _n}}}}\right)}T_s}\\	\text{s.t.}\quad 
	& \eqref{allocatedtime}, \eqref{q0qF}, \eqref{timeslot}, \eqref{delta}, \eqref{throughputconstraint2}, \eqref{Ank},\notag\\
	& {\big\| {{\mathbf{v}_e^{(l)}[n]}- {\mathbf{v}_w^{(l)}}[n]} \big\|^2}+2{\big( {{\mathbf{v}_e^{(l)}}[n]-{\mathbf{v}_w^{(l)}}[n]} \big)^T}\big({\mathbf{v}_e}[n]-{\mathbf{v}_w}[n]-{\mathbf{v}_e^{(l)}}[n]+{\mathbf{v}_w^{(l)}}[n]\big)\ge\delta _n^2, n\in\mathcal{N}\setminus\{0\},\label{localvw}\\
	&\mathbf{q}[n+1] \approx \mathbf{q}[n] + \frac{1}{2}({{\mathbf{v}_e}[n+1]+{\mathbf{v}_e}[n]}){T_s}, n\in\mathcal{N}, \label{q3}\\
	&{\mathbf{v}_e[0]}-{\mathbf{v}_w[0]} = {\mathbf{v}_0}, {\mathbf{v}_e[N+1]}-{\mathbf{v}_w[N+1]}= {\mathbf{v}_F},\label{v0vF2}\\ 
	&\left\| {{\mathbf{v}_e}[n]-{\mathbf{v}_w}[n]} \right\| \le {V_{\max }}, n\in\mathcal{N}\setminus\{0\},\label{vconstraint2}\\
	&\left\| \frac{{{\mathbf{v}_e}[n+1]-{\mathbf{v}_w}[n+1]-{\mathbf{v}_e}[n] + {\mathbf{v}_w}[n]}}{{{T_s}}} \right\| \le {a_{\max }}, n\in\mathcal{N},\label{a3}
\end{align}%
\end{small}%
where ${{\mathbf{v}}[n]}={{\mathbf{v}_e}[n]}\!-\!{{\mathbf{v}}_w[n]}$, and ${\mathbf{v}_e^{(l)}[n]}$ (or ${\mathbf{v}_w^{(l)}[n]}$) is the local point of ground velocity (or wind velocity) in each iteration.
Then, in order to improve the robustness against wind variations, we transform the wind-related constraints \eqref{localvw} and \eqref{v0vF2} - \eqref{a3} into expected inequality constraints. In particular, the equality constraints in \eqref{v0vF2} are transformed into the form of $\mathbb{E}_{\{\mathbf{v}_w\}} \left(\left\| {{h_j}(\mathbf{x})} \right\|\right) \le \varepsilon$, where $\varepsilon$ is a small tolerance value\cite{Equivalent}.
By taking the wind statistics into account, such transformation increases the robustness and flexibility of the obtained offline solution, which, when applied in real time, is less likely to cause violation of the flight and communication constraints.
As a result, problem (P3.1) can be reformulated as

\begin{small}
\begin{align}
	\mathrm{(P3.2)}:& \underset{
		\begin{subarray}{c}
			\{ {\mathbf{q}[n]}\} ,\{ {\mathbf{v}_e[n]}\},\{ {\tau_{k}[n]}\},\\
			\{ {A_{k}[n]}\},\{ {\delta _n}\}\notag\\
		\end{subarray}
	}{\min}{\sum\limits_{n=1}^{N}{\left({{w_1}{{\left\| {{\mathbf{v}_e}[n]-\mathbb{E}({\mathbf{v}_w})} \right\|}^3}+\frac{{{w_2}}}{{{\delta _n}}}+\frac{{{w_2}{{\left\| {{\mathbf{v}_e}[n+1]-{\mathbf{v}_e}[n]} \right\|}^2}}}{{{{\left( {g{T_s}} \right)}^2}{\delta _n}}}}\right)}T_s}\\	
	\text{s.t.}\quad 
	& \eqref{allocatedtime}, \eqref{q0qF}, \eqref{timeslot}, \eqref{delta}, \eqref{throughputconstraint2}, \eqref{Ank},\eqref{q3},\notag\\
	& \mathbb{E}\left( {\big\| {{\mathbf{v}_e^{(l)}}[n]- \mathbb{E}(\mathbf{v}_w)} \big\|^2}+2{\big( {{\mathbf{v}_e^{(l)}}[n]-\mathbb{E}(\mathbf{v}_w)} \big)^T}\big({\mathbf{v}_e}[n]-{\mathbf{v}_w}[n]-{\mathbf{v}_e^{(l)}}[n]+\mathbb{E}(\mathbf{v}_w)\big) \right) \ge \delta _n^2, n\in\mathcal{N}\setminus\{0\},\label{vw}\\
	&\mathbb{E} \left( \left\| \frac{{{\mathbf{v}_e}[n+1]-{\mathbf{v}_w}[n+1]-{\mathbf{v}_e}[n] + {\mathbf{v}_w}[n]}}{{{T_s}}} \right\| \right) \le {a_{\max }}, n\in\mathcal{N},\label{eamax}\\
	&\mathbb{E} \left( \left\| {\mathbf{v}_e[0]}-{\mathbf{v}_w}[0]-{\mathbf{v}_0} \right\| \right) \le \varepsilon_1, \mathbb{E} \left( \left\| {\mathbf{v}_e[N+1]}-{\mathbf{v}_w}[N+1]-{\mathbf{v}_F} \right\| \right) \le \varepsilon_2,\\ 
	&\mathbb{E}\left( {\left\| {{\mathbf{v}_e}\left[ n \right] - {\mathbf{v}_w}\left[ n \right]} \right\|} \right) \le {V_{\max }}, n\in\mathcal{N}\setminus\{0\},\label{eqmax}
\end{align}%
\end{small}%
where $\varepsilon_1$ and $\varepsilon_2$ are small tolerance values. Since expectations preserve convexity, the expected inequality constraints \eqref{vw} - \eqref{eqmax} are still convex.
In addition, for simplicity, we choose the local point with ${\mathbf{v}_w^{(l)}[n]}=\mathbb{E}(\mathbf{v}_w)$ in \eqref{vw}.
Similarly, we substitute ${{\mathbf{v}}[n]}={{\mathbf{v}_e}[n]}\!-\!\mathbb{E}(\mathbf{v}_w)$ into the objective function of (P3.2), which, by Jensen's inequality, provides a lower bound for the mean of cost function in problem (P3.1).
As a result, problem (P3.2) is a convex stochastic program (CSP).

To solve the expectation constraints in problem (P3.2), we use simple Monte Carlo evaluation, \emph{i.e}, $\mathbb{E}f(x,\omega ) \approx \frac{1}{{{S}}}\sum\limits_{i = 1}^{{S}} {f(x,{\omega _i})}$, where $f$ is assumed to be convex in the optimization variable $x$ for each value of the random variable $\omega$, and $\omega_i$, $i = 1, \ldots, S$ are samples of $\omega$; this approximation is referred to as the \emph{sample average approximation} (SAA) in the SP literature, and methods that use it are often referred to as \emph{scenario-based} methods. Furthermore, both the optimal value and optimal set of a SAA converge almost surely to the optimal value and optimal set of the true problem (P3.2)\cite{2015Disciplined}.
Based on the above, all the expectation constraints in problem (P3.2) can be re-written in the approximate form $\mathbb{E}f({\mathbf{v}_e[n]},\mathbf{v}_w ) \approx \frac{1}{{{S}}}\sum\limits_{i = 1}^{{S}} {f({\mathbf{v}_e[n]},\mathbf{v}_{w,i} )}$. For example, the inequality \eqref{eqmax} is approximated as $\frac{1}{{{S}}}\sum\limits_{i = 1}^{{S}}\left( {\left\| {{\mathbf{v}_e}\left[ n \right]\!-\!{\mathbf{v}_{w,i}}\left[ n \right]} \right\|} \right) \le {V_{\max }}, n\in\mathcal{N}\setminus\{0\}$.

Next, we discuss the computational complexity for solving problem (P3.2) using the standard interior-point method, based on the complexity analysis method in \cite{complexity}.
Specifically, the problem involves $2N\!+\!3$ second-order cone (SOC) constraints of size $2S$ in \eqref{eamax}-\eqref{eqmax}, $KN\!+\!K\!+\!N$ SOC constraints of size $2$ in \eqref{Ank} and \eqref{vw}, 1 SOC constraints of size $2N$ in the first term of the objective function, $KN\!+\!2N\!+\!K\!+\!3$ linear matrix inequality (LMI) constraints of size $2$ in \eqref{q0qF}, \eqref{Ank}, \eqref{q3} and the second term of the objective function, $KN\!+\!2N\!+\!2K\!+\!1$ LMI constraints of size $1$ in \eqref{allocatedtime}, \eqref{timeslot}, \eqref{delta} and \eqref{throughputconstraint2}, and $N$ LMI constraints of size $3$ in the third term of the objective function,
while the total number of optimization variables is $u\!\triangleq\!2KN\!+\!3N$. Therefore, based on the analysis in \cite{complexity}, the number of iterations required for obtaining the optimal solution scales as $\sqrt {5KN\!+\!15N\!+\!6K\!+\!15}$, while the required complexity in each iteration is given by $u[ \left( {9KN\!+\!45N\!+\!10K\!+\!25} \right)\!+u\left( {5KN\!+\!19N\!+\!6K\!+\!13} \right)\!+\!\left( {4(2N{S^2}\!+\!3{S^2}\!+\!{N^2}\!+\!KN\!+\!N\!+\!K)} \right)\!+\! {u^2} ]$. Then the required complexity for solving problem (P3.2) scales as $O(a_1(KN)^{3.5}\!+\!a_2K^{1.5}S^2N^{2.5})$, with coefficients $a_1$ and $a_2$. Since the number of time slots $N$ is generally greater than the number of samples $S$, the complexity is still dominated by $N$, and thus the SAA involves only moderate increase in complexity. 
Finally, note that the solution obtained in Section \ref{fixed wind} under fixed wind (with ${\mathbf{v}_w[n]}=\mathbb{E}(\mathbf{v}_w)$, $\forall n$) provides a good initial solution for the SP-based solution in this section, which also helps to speed up the computation in practice.

\section{Online Adaptation to Real-Time Wind Effect}\label{online}
The above offline design based on fixed-wind assumption in Section \ref{fixed wind} or SP method in Section \ref{offlinedesign}, strives to minimize the overall energy consumption by jointly optimizing the UAV trajectory and communication time scheduling across the whole mission period subject to statistical wind effect. Despite being statistically favorable, the obtained solution may not be optimal or even feasible when faced with real-time wind. This issue is addressed in this section by designing the online policy that adjusts the UAV's flying speed along the offline optimized UAV path as well as its communication time scheduling in adaption to the real-time wind velocity and the buoys’ individual amount of data received accumulatively. 
Specifically, the offline solution obtained in Section \ref{fixed wind} or \ref{offlinedesign}, denoted as $\{ {\mathbf{q}^{\textup{off}}[n]}\}$, $\{ {\mathbf{v}_e^{\textup{off}}[n]}\}$ and $\{ {\tau_{k}^{\textup{off}}[n]}\}$, serves as a reference for real-time operation. As such, the online adaptation should not deviate too much from the offline reference in order to minimize the overall energy consumption.
On the other hand, due to the randomness of real-time wind velocity, it could happen that the planned UAV trajectory and airspeed as well as the scheduled data collection may not be executed exactly during online operation.
Therefore, for each current time slot $n$, we introduce slack variables to relax the flight and communication constraints in order to make the online adaptation feasible and more flexible.

In particular, regarding data collection, the amount of collected data in bits at the UAV from each buoy $k$ over the $n$-th time slot is given by $\tau_{k}[n]{R_k}[n-1]$, and thus the amount of data collected accumulatively up to the beginning of each time slot $n$, denoted by ${Q_{k,n}^{\textup{acc}}}$, evolves as
\begin{equation}
{Q_{k,1}^{\textup{acc}}} \triangleq 0, {Q_{k,n}^{\textup{acc}}} \triangleq \sum_{m = 1}^{n - 1} {{\tau_{k}[m]}} {R_k}[m-1], n > 1, \forall k.
\end{equation}
Furthermore, the amount of data to be collected over all the subsequent time slots $m\!=\!n\!+\!1, \cdots, N\!+\!1$, can be obtained from the offline phase, which is denoted by ${Q_{k,n}^{\textup{off}}}\! \triangleq\!\sum\limits_{m = n+1}^{N+1}\!{{\tau_k^{\textup{off}}[m]}} {R_k^{\textup{off}}}[m\!-\!1],\forall k$, where ${R_k^{\textup{off}}}[m]$ is the communication rate along the offline time slot $m$.
To satisfy the overall data volume requirement ${\bar Q_{{k}}}$, the throughput constraint in \eqref{throughputconstraint} for the current time slot $n$ is re-written as
\begin{equation}
	B{{\tau}_{k}[n]} {\log _2}\bigg(1 + \frac{{{\gamma _0}}}{{{H^2} + {{\left\| {{\mathbf{q}[n\!-\!1]}-{\mathbf{b}_k}} \right\|}^2}}}\bigg)  \ge {\bar Q_{{k}}}-{Q_{k,n}^{\textup{acc}}}-{Q_{k,n}^{\textup{off}}}-\zeta,\forall k, \label{throughputconstraint3}
\end{equation}
where $\mathbf{q}[n\!-\!1]$ is the previous waypoint traversed online; and the data volume to be collected in the current time slot $n$ is determined by the accumulated volume ${Q_{k,n}^{\textup{acc}}}$ so far and the volume ${Q_{k,n}^{\textup{off}}}$ to be collected in the subsequent time slots. 
Moreover, 
a slack variable $\zeta\geq 0$ is introduced to relax the data collection constraint in order to achieve a feasible and more flexible solution. The larger $\xi$ is, the easier it is to satisfy the data collection constraint in the current time slot, which, however, has a larger impact on the completion of subsequent collection tasks. In order to limit the scope of relaxation, a cost term $w_3\zeta$ is added to the objective function.
As a result, the online adaptation problem at each time slot $n$ can be modified from that for the offline design as follows, i.e.,
	\begin{align}
	\mathrm{(P4.1)}:& \underset{
		\begin{subarray}{c}
		\mathbf{v}_e[n]; \tau_k[n], k\in\mathcal{K}; \zeta\ge 0\notag\\
		\end{subarray}
	}{\min}{{{{w_1}{{\left\| {{\mathbf{v}[n\!-\!1]}} \right\|}^3}T_s+\frac{{{w_2}}}{{\left\| {{\mathbf{v}[n\!-\!1]}} \right\|}}T_s+\frac{{{w_2}{{\left\| {{\mathbf{v}[n]}-{\mathbf{v}[n\!-\!1]}} \right\|}^2}}}{{{{g}^2}}{\left\| {\mathbf{v}[n\!-\!1]} \right\|}T_s}}+w_3\zeta}}\\ \text{s.t.}\quad &\eqref{throughputconstraint3},\notag\\
	&\textstyle{\sum}_{k = 1}^K {\tau}_{k}[n]\le T_s; \quad {\tau_{k}[n]} \ge 0,\forall k,\label{timeCons}\\
	&\left\| \mathbf{q}[n-1] + \frac{1}{2}({\mathbf{v}_{e}[n]+\mathbf{v}_e[n-1]}){T_s} - \mathbf{q}^{\textup{off}}[n] \right\| \le \xi_q, \label{q4}\\
	&\left\| \mathbf{v}_e[n]-\mathbf{v}_e^{\textup{off}}[n]\right\| \le \xi_v,\label{veoffcons}\\
	&\left\| \frac{{\mathbf{v}_{e}[n]-{\mathbf{v}_w}[n]-\mathbf{v}_e[n\!-\!1] + {\mathbf{v}_w}[n\!-\!1]}}{{{T_s}}} \right\| \le {a_{\max }},\label{amx}\\
	&{V_{\min}} \le \left\| {{\mathbf{v}_{e}[n]}-{\mathbf{v}_w}[n]} \right\| \le {V_{\max }}, \label{vemaxmin}	
	\end{align}
where ${\mathbf{v}[n]}={\mathbf{v}_{e}[n]}-{\mathbf{v}_w[n]}$, and $\xi_q$ and $\xi_v$ are two small tolerance values to restrict the extend of deviation from the offline trajectory and ground velocity, respectively.
Note that we record $\mathbf{q}[n-1]$, $\mathbf{v}_e[n\!-\!1]$ and ${\mathbf{v}_w}[n\!-\!1]$ from the previous time slot, and also measure the wind velocity ${\mathbf{v}_w[n]}$ for the current time slot.
Therefore, it can be verified that the objective function and the constraints except \eqref{vemaxmin} are convex in the optimization variables $\mathbf{v_e}[n]$, $\zeta$ and $\tau_k[n], k\in\mathcal{K}$.
To deal with the non-convex constraint of $\left\| {{\mathbf{v}_{e}[n]}-{\mathbf{v}_w}[n]} \right\|\ge {V_{\min}}$, we follow similarly the transforms as in \eqref{vappro} and \eqref{vapproximation}, i.e., we introduce a slack variable $\delta_n$ and employ the first-order Taylor expansion at the sub-optimal local point of airspeed ${\mathbf{v}'[n]}\triangleq{\mathbf{v}_e^{\textup{off}}}[n]-\mathbb{E}(\mathbf{v}_w)$.
As a result, problem (P4.1) is reformulated as
	\begin{align}
	\mathrm{(P4.2)}:& \underset{
		\begin{subarray}{c}
		\mathbf{v_e}[n]; \tau_k[n], k\in\mathcal{K}; \\
		{\delta _n};\zeta\ge 0\notag\\
		\end{subarray}
	}{\min}{{{{w_1}{{\left\| {{\mathbf{v}[n\!-\!1]}} \right\|}^3}T_s+\frac{{{w_2}}}{{\left\| {{\mathbf{v}[n\!-\!1]}} \right\|}}T_s+\frac{{{w_2}{{\left\| {{\mathbf{v}[n]}-{\mathbf{v}[n\!-\!1]}} \right\|}^2}}}{{{{g}^2}}{\left\| {\mathbf{v}[n\!-\!1]} \right\|}T_s}}+w_3\zeta}}\\ \text{s.t.}\quad &\eqref{throughputconstraint3}-\eqref{amx},\notag\\
	&{\left\| {{\mathbf{v}_e^{\textup{off}}}[n]-\mathbb{E}(\mathbf{v}_w)} \right\|^2}+2{\left( {{\mathbf{v}_e^{\textup{off}}}[n]-\mathbb{E}(\mathbf{v}_w)}\right)^T}({\mathbf{v}_e}[n]-{\mathbf{v}_w}[n]-{\mathbf{v}_e^{\textup{off}}}[n]+\mathbb{E}(\mathbf{v}_w))\ge\delta _n^2,\\
&{\delta _n} \ge {V_{\min}}, \quad \left\| {{\mathbf{v}_{e}[n]}-{\mathbf{v}_w}[n]} \right\| \le {V_{\max }},
	\end{align}
which can be verified to be a convex quadratically constrained quadratic program (QCQP) with only $K+3$ variables, and thus can be solved efficiently.
In particular, based on similar analysis as in Sections \ref{fixed wind} and \ref{offlinedesign}, the complexity of solving problem (P4.2) using the interior-point method is reduced to $O(K^{3.5})$, which does not require global optimization involving all $N$ time slots. Therefore, problem (P4.2) can be solved efficiently, rendering the online adaptation amenable to practical implementation. 

Finally, note that refining the trajectory, airspeed and time scheduling in each time slot helps ensure the feasibility of the reference solution obtained in the offline phase. Specifically, at each time slot online, the UAV acquires the real-time wind velocity and then adjusts the airspeed to achieve lower energy consumption. Moreover, by relaxing the constraints on the next trajectory point and ground speed in \eqref{q4} and \eqref{veoffcons}, as well as the amount of data volume to be collected in \eqref{throughputconstraint3}, the optimization space is expanded, and online adaptation becomes more flexible and robust.


\section{Numerical Results}\label{simulation}
	
This section provides numerical results to validate the proposed design. The following parameters are used if not mentioned otherwise: $H=100$ m, $B=1$ MHz, ${\gamma _{\rm{0}}}=70$ dB, $V_\textrm{max}=50$ m/s, $V_s=3$ m/s, $a_\textrm{max}=5$ m/s$^2$, $w_1=9.26 \times {10^{ - 4}}$, $w_2=2250$\cite{ZengEnergy}, $w_3=100$ , $\rho_c=0.5$, $S=100$, $\varepsilon_1=\varepsilon_2=1$ m/s, $\xi_q=3$ m and $\xi_v=0.2$ m/s.
We first consider the single-buoy scenario and investigate two cases including the chain-like flight (explained later) and cyclical flight, which help illustrate the effect of wind on the data collection task as well as energy consumption, for scenarios with/without wind, or under different wind speed/direction/variance.
Then we extend to the multi-buoy case, and investigate the situations where the distributed buoys are far from/close to each other.

\subsection{Single Buoy Setup}
\subsubsection{Fixed Wind}
\begin{figure}[h]
		\centering
		\includegraphics[height=0.38\linewidth,width=0.50\linewidth]{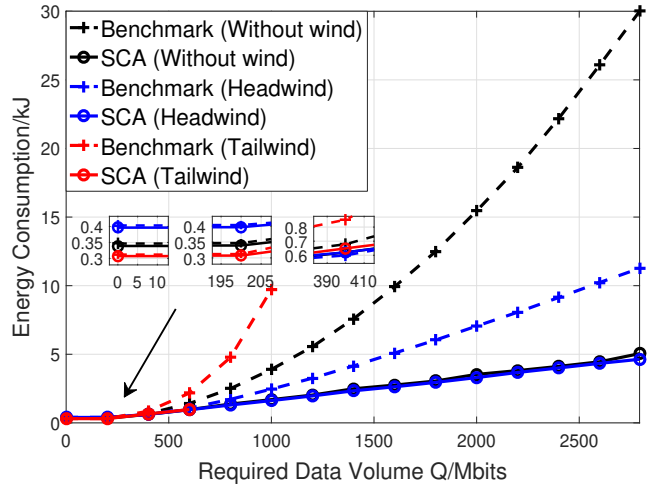}
		\caption{Energy consumption versus data volume requirement under different wind conditions.\vspace{-1ex}}
		\label{QVE}
\end{figure}

\paragraph{Chain-Like Flight}\label{SectionChain}
Assume $\mathbf{q}_{0}={[-600,0]^T}$ m and $\mathbf{q}_{F}={[600,0]^T}$ m with the single buoy at the origin. In this case, the initial point $\mathbf{q}_{0}$, the buoy and the final point $\mathbf{q}_{F}$ make up a chain-like topology, and hence the UAV's flight is termed as the \textit{chain-like flight}.
The UAV's initial trajectory is set to be the direct path from $\mathbf{q}_\textrm{0}$ to $\mathbf{q}_{F}$.
Assume that the wind speed is $V_w=5$ m/s, which either blows from $\mathbf{q}_{0}$ to $\mathbf{q}_{F}$ (i.e., tailwind) or reversely (i.e., headwind).
We use SCA to optimize the UAV's trajectory and airspeed, and also compare with the benchmark scheme where the UAV flies along a straight line at a constant but optimized airspeed. 
The resulted energy consumption for collecting different data volume $Q$ is shown in Fig. \ref{QVE}, under different wind conditions.
	
For the benchmark scheme,
in the case with small $Q$ (e.g., $Q\leq 200$ Mbits), it is observed that the tailwind helps achieve lower energy consumption compared with the headwind or no-wind case.
This is because collecting small data volume $Q$ can be done easily, and the main task of the UAV is to fly from $\mathbf{q}_{0}$ to $\mathbf{q}_{F}$, for which the tailwind helps.
On the other hand, as $Q$ increases (e.g., $Q\geq 400$ Mbits), the tailwind consumes the most energy while the headwind consumes the least energy, which might be \textit{against the intuition} that the tailwind adds thrust to the UAV and hence should help save energy.
The underlying reason is that, to collect more data, the UAV has to slow down to allow more time to communicate with the buoy, and hence might not fly at the most energy-saving airspeed\footnote{Based on \eqref{energyconsumption}, there exists a most energy-saving airspeed, since the UAV consumes much energy at both high airspeed and also low airspeed (in order to keep the UAV aloft).}.
Note that due to the limited communication time, the UAV might not be able to collect too large data volume (e.g., $Q> 1000$ Mbits) in the tailwind case.
	
\begin{figure}[h]
		\centering
		{
			\label{a} 
			\includegraphics[height=0.30\linewidth,width=0.37\linewidth]{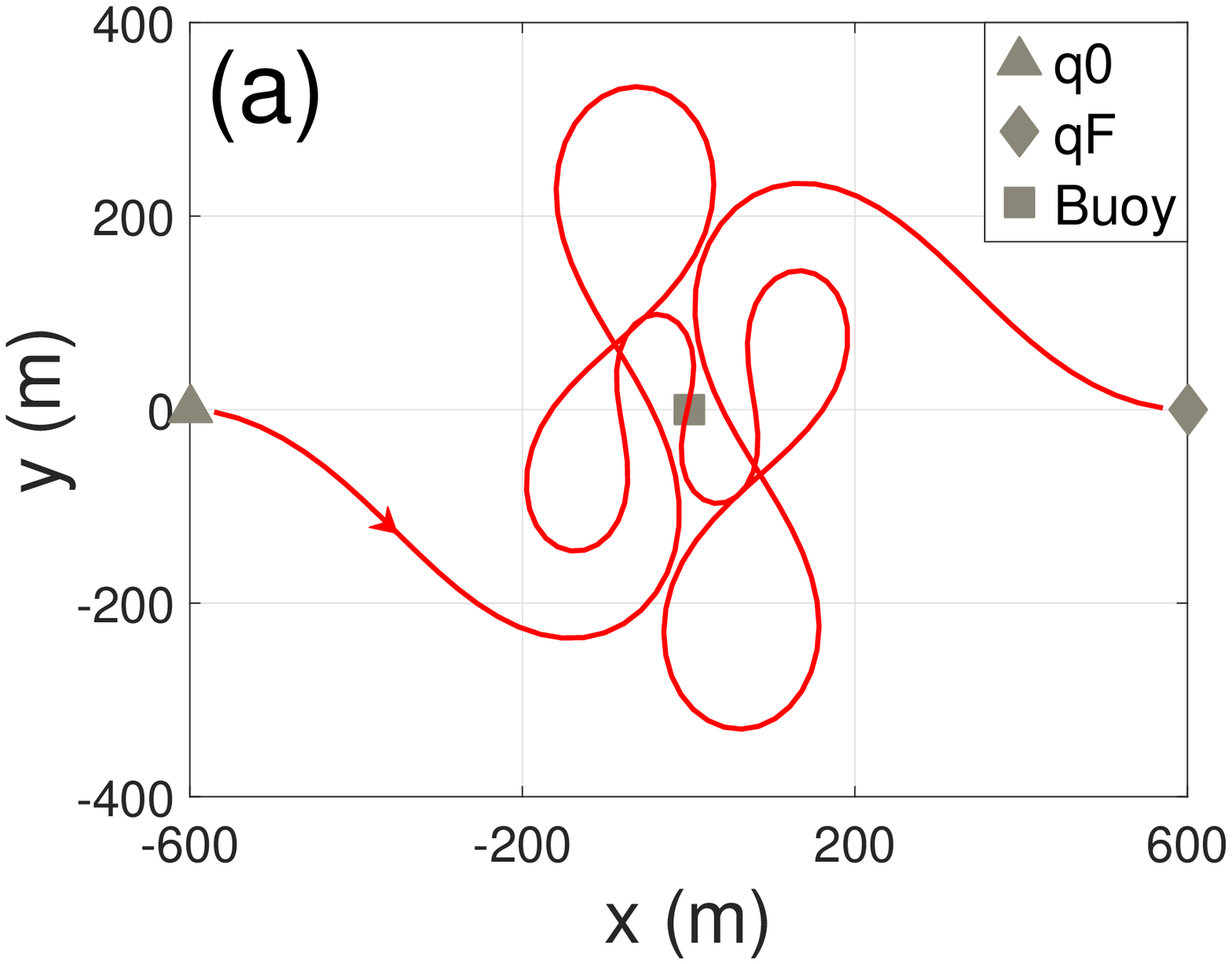}}
		{
			\label{b} 
			\includegraphics[height=0.30\linewidth,width=0.37\linewidth]{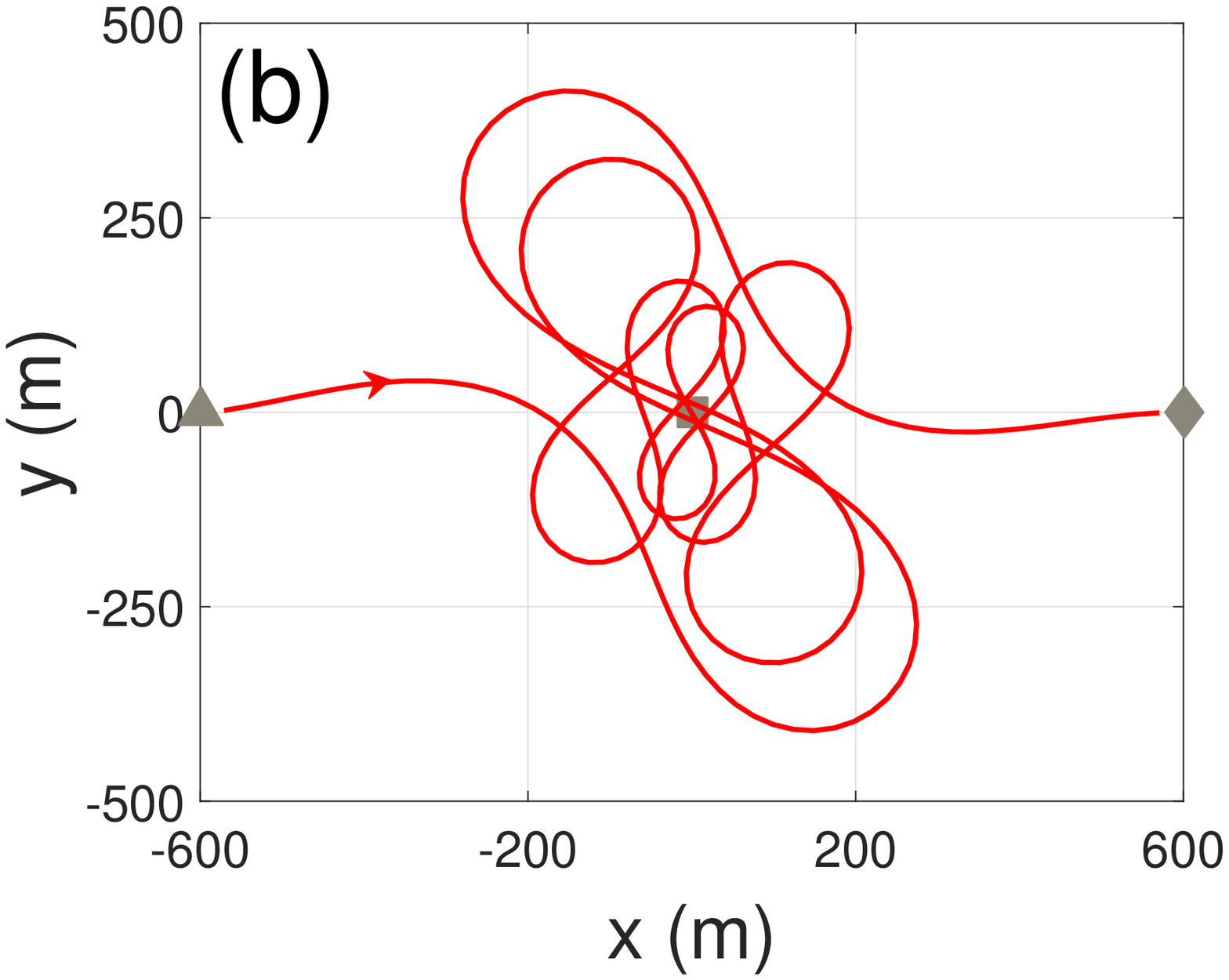}}
		\caption{The SCA-based UAV trajectory for the chain-like flight without wind, under (a) $Q=1400$ Mbits and (b) $Q=2200$ Mbits.\vspace{-1ex}}
		\label{CurveTrajectory} 
\end{figure}

In comparison, regardless of the wind conditions, our proposed SCA-based scheme is able to adapt to the wind and achieve significant energy savings compared with the benchmark scheme.
However, it is worth noting that as the required data volume increases, the computational complexity and the resulted trajectory complexity of the SCA-based solution both increase dramatically.
Two examples of the UAV trajectory under different $Q$ are shown in Fig. \ref{CurveTrajectory}. 
This thus motivates our cyclical trajectory design to achieve lower energy consumption with less computational time and simpler UAV trajectory.
	
\begin{figure}[h]
		\centering
		\includegraphics[height=0.38\linewidth,width=0.50\linewidth]{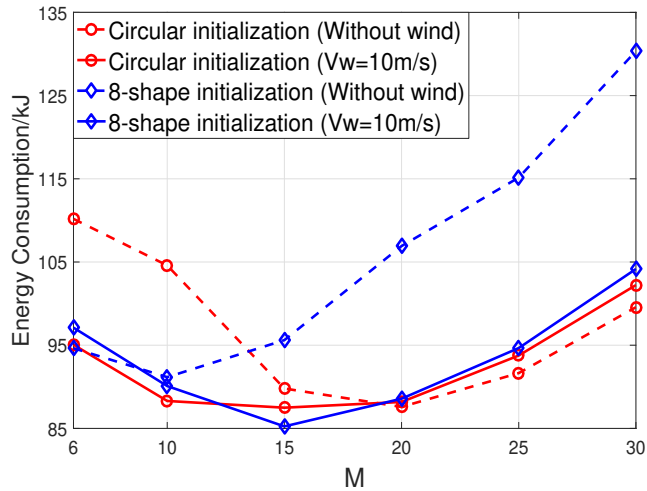}
		\caption{Total energy consumption of all $M$ laps under different trajectory initializations, with or without wind.\vspace{-1ex}}
		\label{MVE}
\end{figure}

\paragraph{Cyclical Trajectory}\label{SectionCyclical}
Consider a large data volume, e.g., $Q=6$ Gbits.
For our considered single-buoy setup, the proper amount of data volume $Q_0=Q/M$ for each lap should be around 200 to 1000 Mbits, based on the observations in Section \ref{SectionChain}.
Therefore, we choose the number of laps $M$ in the range of 6 to 30. 
Based on our proposed cyclical trajectory optimization in Algorithm \ref{CyclicalTrajectory}, the resulted total energy consumption of all $M$ laps is shown in Fig. \ref{MVE}, under different trajectory initializations and wind conditions.
It is observed that with our proposed cyclical trajectory optimization, both the optimized circular and 8-shape trajectories can proactively exploit the wind to reduce the energy consumption under a certain $Q_0$ per lap, compared with the case without wind.
In particular, the 8-shape trajectory can even make better use of the wind and outperform the circular trajectory in some cases (e.g., under $Q_0=400$ Mbits and $V_w=10$ m/s).
We provide more detailed discussions next.
	
\textbf{Without Wind}: In this case, the circular trajectory performs better than the 8-shape trajectory when $M\geq 15$, where the minimum total energy consumption occurs at $M=20$ (i.e., $Q_0=300$ Mbits) with the circular trajectory. 
This is because when $Q_0$ is relatively small, the UAV can collect data by flying circularly with a moderate acceleration, without incurring much turning energy as in the 8-shape trajectory.
On the other hand, when $Q_0$ is large, it is more beneficial to adopt the 8-shape flight whose trajectory points are overall closer to the buoy and hence enjoy better communication channel.
	
\begin{figure}[h]
		\centering
		{
			\includegraphics[height=0.3\linewidth,width=0.37\linewidth]{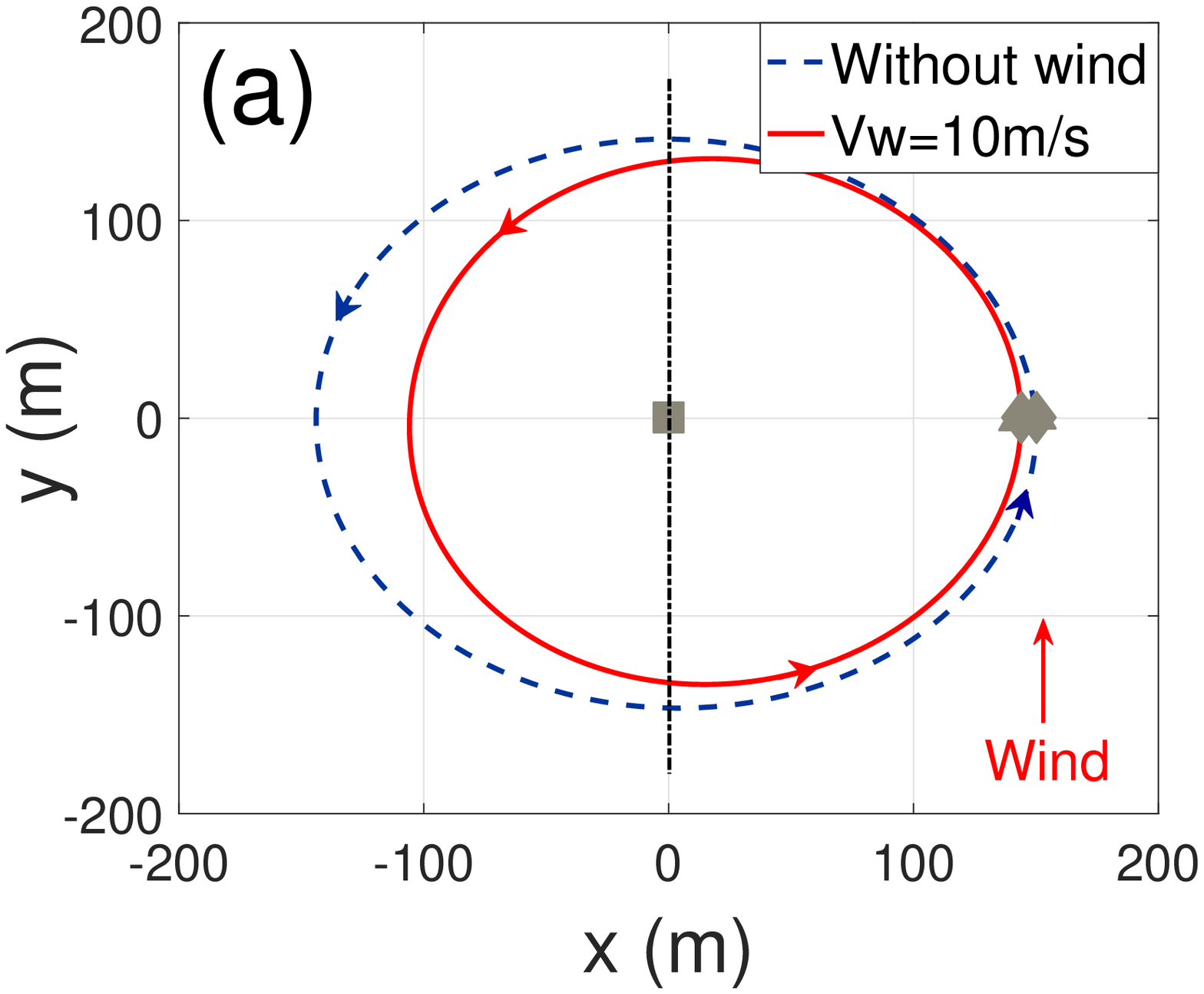}}
		{
			\includegraphics[height=0.3\linewidth,width=0.37\linewidth]{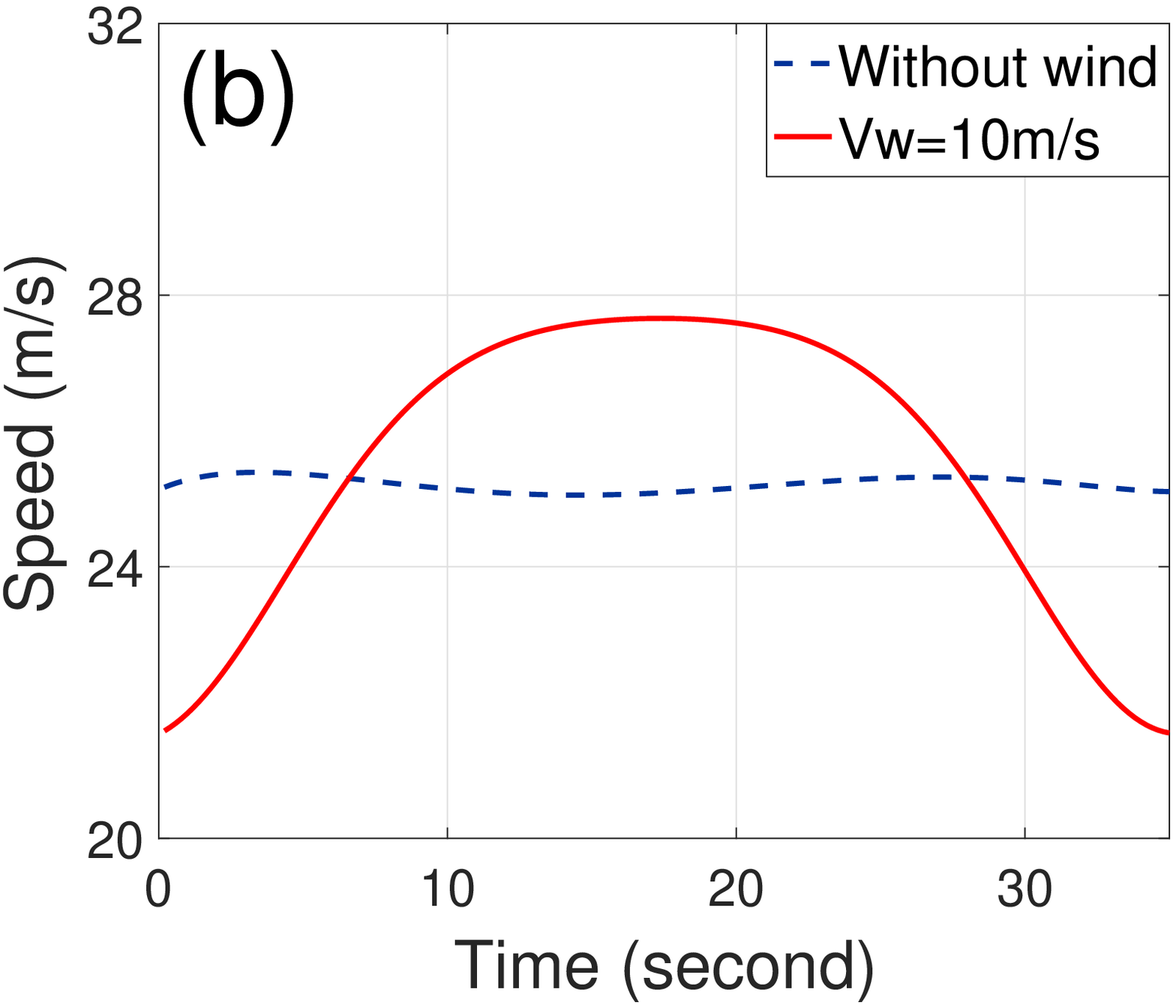}}
		\caption{Optimized UAV (a) trajectory and (b) airspeed for circular trajectory.\vspace{-1ex}}
		\label{Circular} 
\end{figure}

\begin{figure}[h]
		\centering
		{
			\includegraphics[height=0.3\linewidth,width=0.37\linewidth]{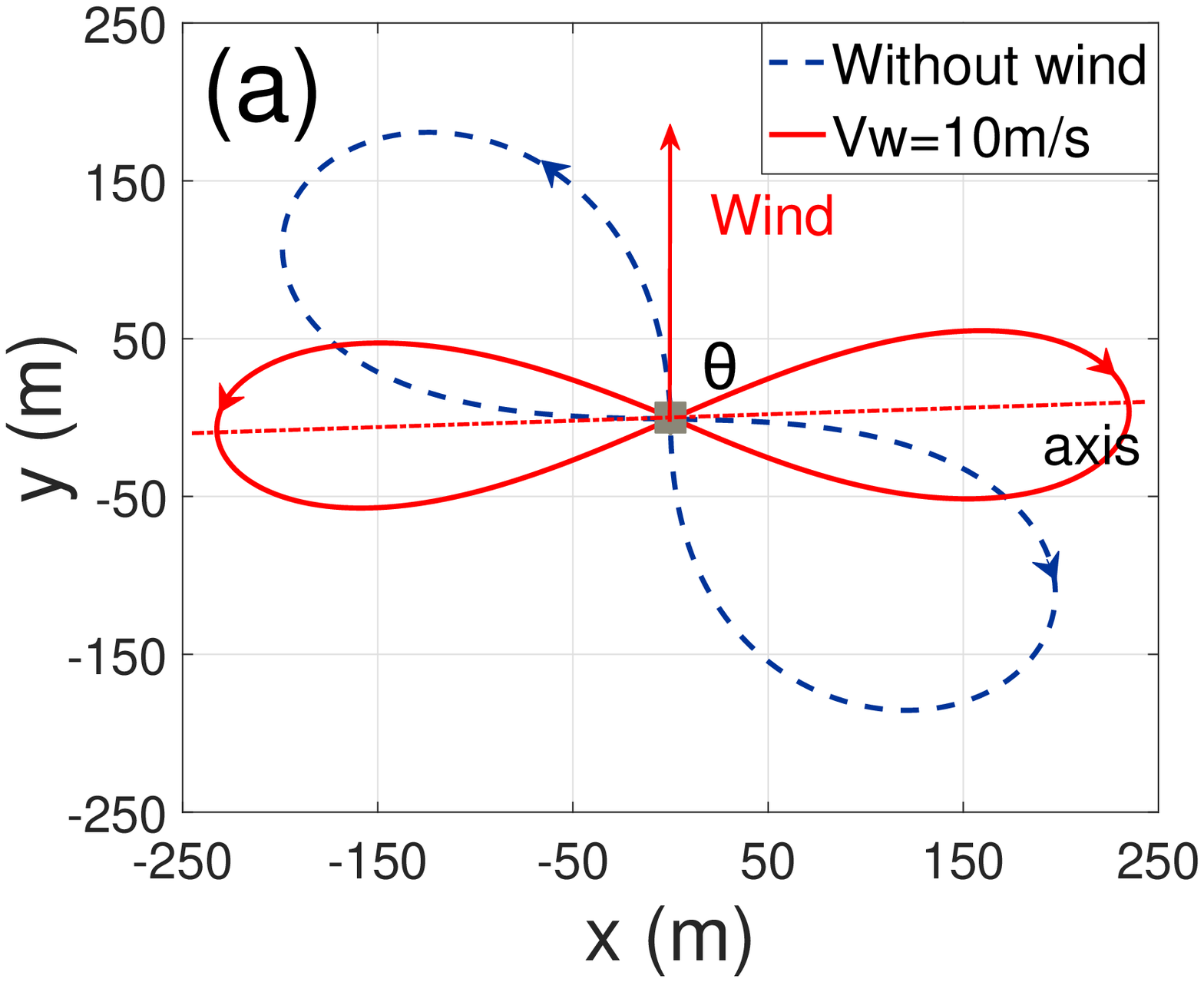}}
		{
			\includegraphics[height=0.3\linewidth,width=0.37\linewidth]{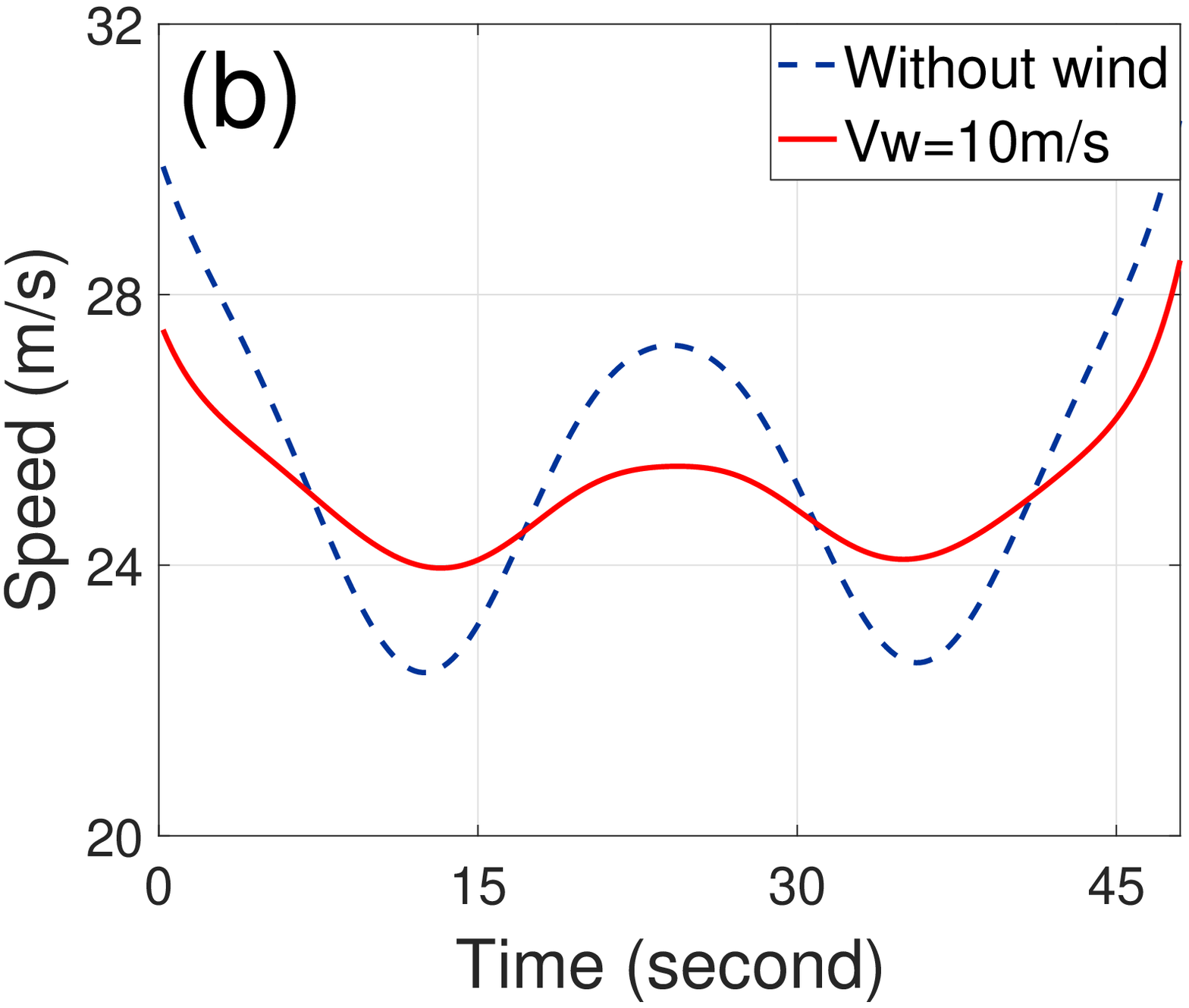}}
		\caption{Optimized UAV (a) trajectory and (b) airspeed for 8-shape trajectory.\vspace{-1ex}}
		\label{8-shape} 
\end{figure}

\textbf{With Wind}: Consider wind blowing from south to north. The optimized UAV trajectory and airspeed for the circular and 8-shape trajectories are shown in Fig. \ref{Circular} and Fig. \ref{8-shape}, respectively.
The circular trajectory is divided into two halves, where the left half experiences headwind and the right half experiences tailwind. The trajectory optimization needs to balance between the UAV's airspeed and angular acceleration in achieving lower energy consumption subject to wind effect.
As for the 8-shape trajectory, it becomes more flatten in wind in order to reduce the overall distance to the buoy.
Moreover, the optimized \textit{orientation} $\theta$, i.e., the angle between the wind and the axis of the 8-shape, is around $90^{\circ}$ as shown in Fig. \ref{8-shape}(a), and hence the UAV experiences headwind in both ends of the 8-shape, whereby the UAV can exploit the wind to slow down and hence reduce the turning energy.

Finally, it is worth noting that the recommended data volume per lap (e.g, $Q_0=300$ Mbits using the optimized circular trajectory for the case without wind, or $Q_0=400$ Mbits using the optimized 8-shape trajectory for the case with wind) serves as a good reference for partitioning arbitrarily large data volume $Q$ into $M=\lceil Q/Q_0\rceil$ laps.
	
\subsubsection{Wind Velocity Change}
\begin{figure}[h]
	\centering
	{
		\includegraphics[height=0.28\linewidth,width=0.32\linewidth]{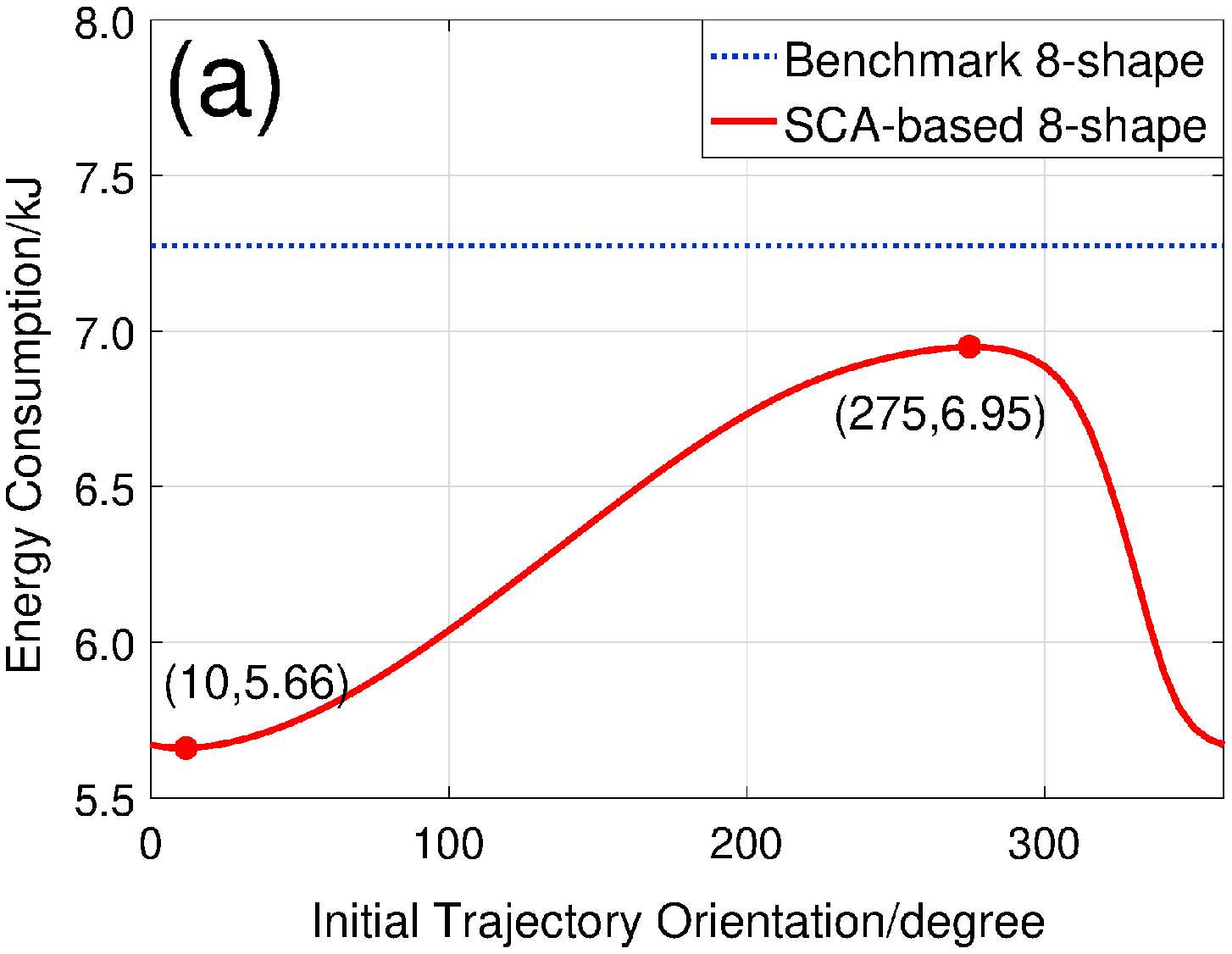}}
	{
		\includegraphics[height=0.28\linewidth,width=0.32\linewidth]{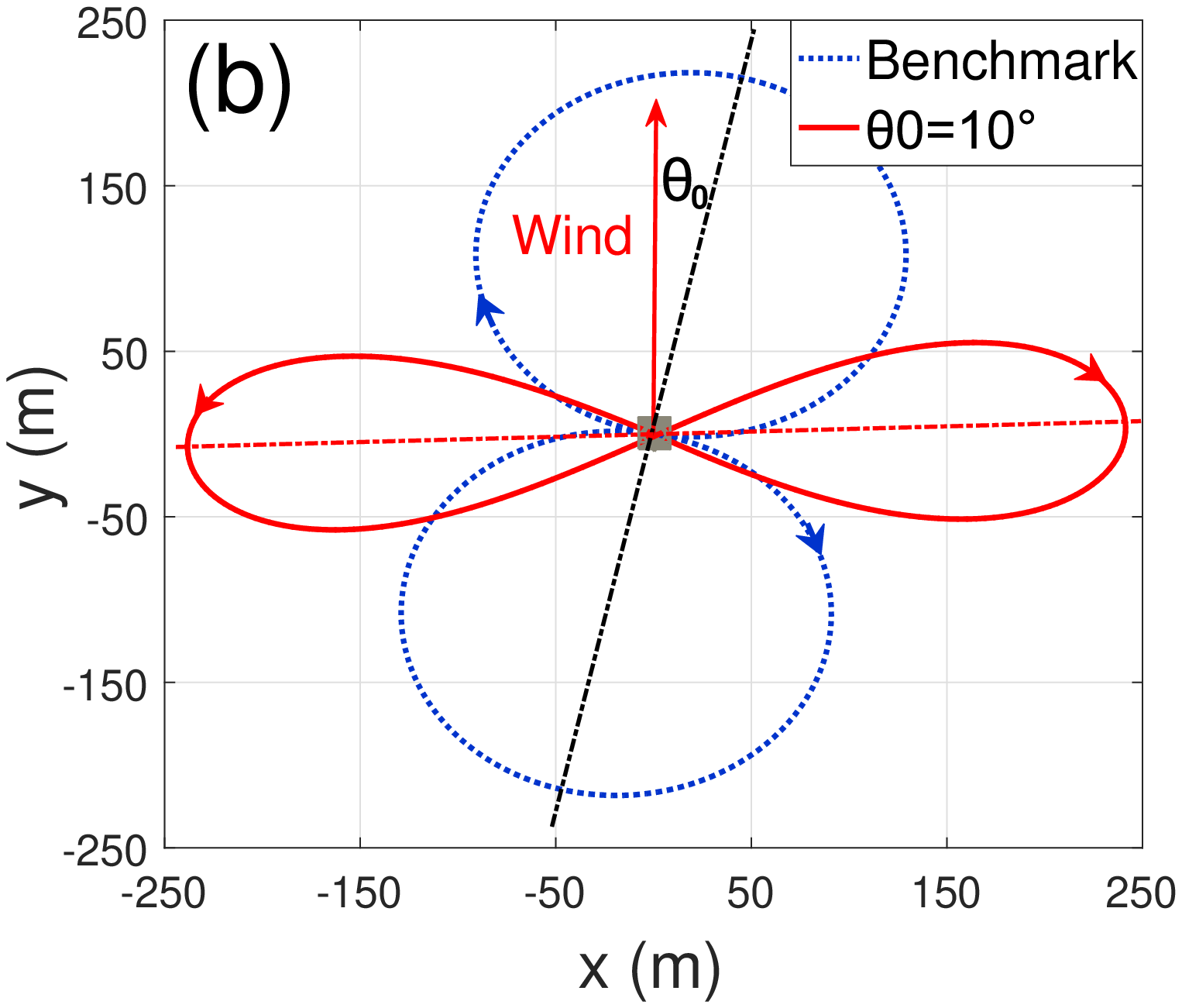}}
	{
		\includegraphics[height=0.28\linewidth,width=0.32\linewidth]{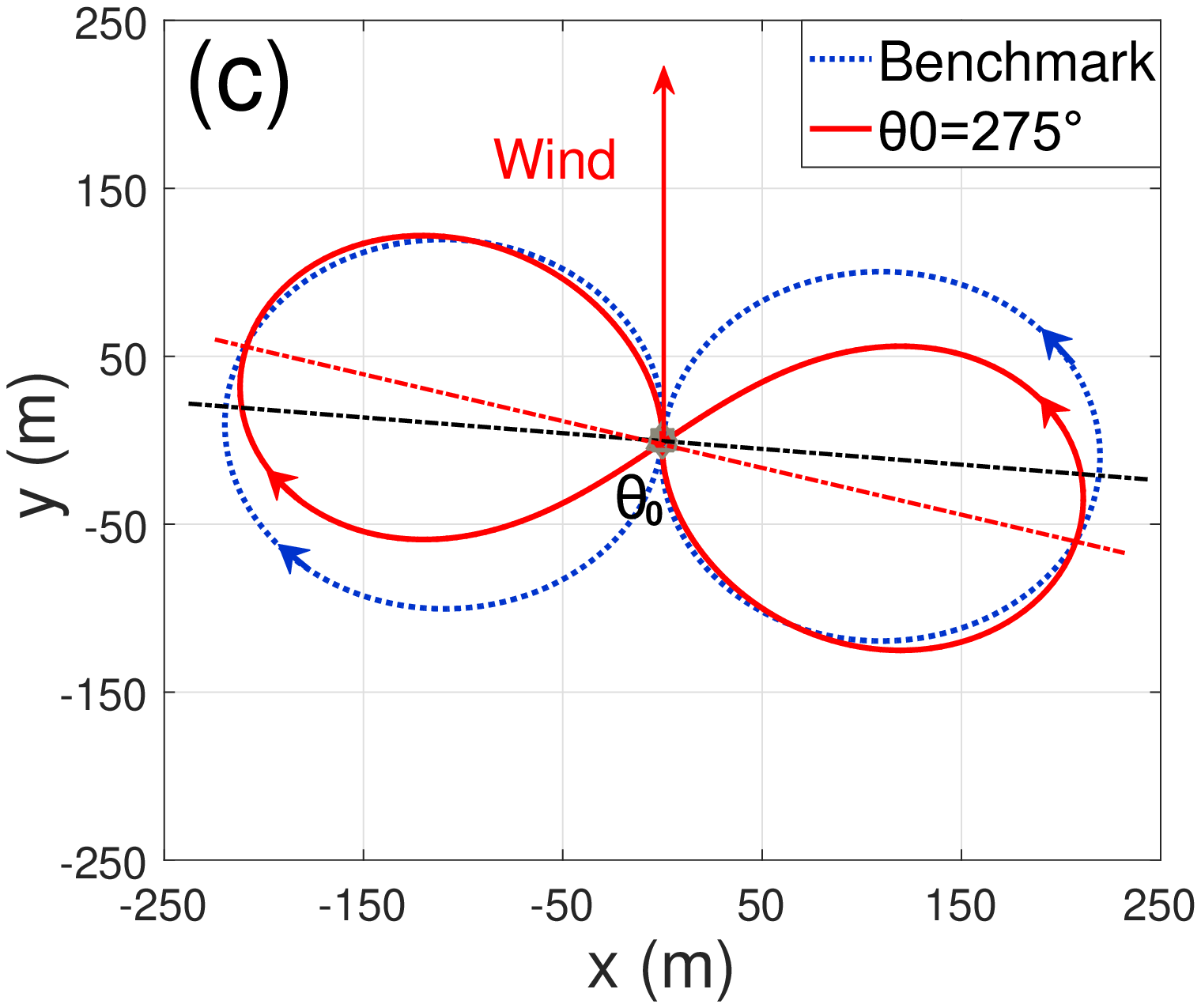}}
	\caption{The effect of different initial trajectory orientation $\theta_0$ on energy consumption. (a) Energy consumption versus trajectory orientation, (b) Benchmark and SCA-based 8-shape trajectory with $\theta_0$=10$^{\circ}$, (c) Benchmark and SCA-based 8-shape trajectory with $\theta_0$=275$^{\circ}$.\vspace{-1ex}}
	\label{orientation} 
\end{figure}

\paragraph{Wind Direction}
For a given wind speed, we first investigate the effect of wind direction on the optimized trajectory and energy consumption. For illustration, we investigate the 8-shape trajectory whose trajectory orientation can be optimized in adaptation to the wind direction and achieve lower energy consumption. Assuming that $V_w=10$m/s and $Q_0=400$ Mbits, Fig. \ref{orientation} shows the energy consumption for the 8-shape trajectory under different trajectory orientation. The benchmark scheme is exact 8-shape trajectory with two circles tangent to each other, just optimizing speed $V$, radius $r$ and period $T_0$. It can be seen that our proposed scheme outperforms the benchmark scheme regardless of the initial orientation, whereby the optimized orientation (e.g., $\theta=10^{\circ}$) in adaptation to the wind direction can further lower the energy consumption. The underlying reason can be similarly explained as in the previous paragraph.

\begin{figure}[h]
	\centering
	\includegraphics[width=0.5\linewidth,  trim=0 0 0 0,clip]{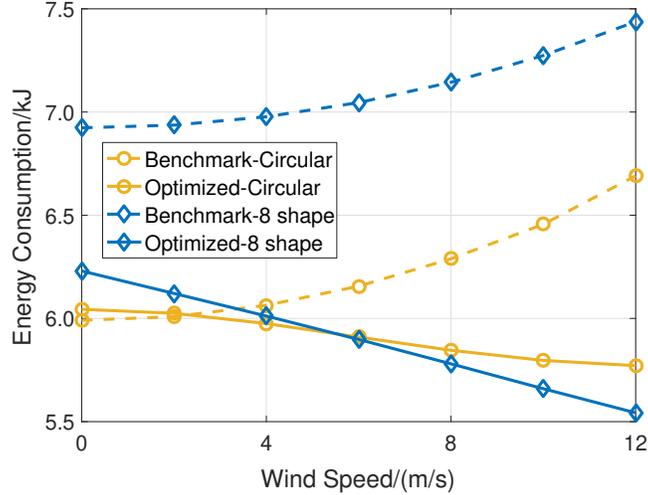}
	\caption{Energy consumption of two trajectory initializations under different wind speed.\vspace{-1ex}}
	\label{Wind speed effect}
\end{figure}

\paragraph{Wind Speed}
Consider $Q_0=400$ Mbits and a given wind direction from south to north. The energy consumption of two trajectory initializations under different wind speed are shown in Fig. \ref{Wind speed effect}. For the benchmark schemes (exact circular/8-shape, just optimizing $V$, $r$ and $T_0$), it is observed that the energy consumption increases with the wind speed. On the other hand, our proposed scheme optimizes the trajectory in adaption to the wind, and surprisingly, even reduces more energy consumption at higher wind speed in this example setup. In particular, due to the anisotropy of the 8-shape trajectory, it can make better use of wind to reduce energy consumption, which is consistent with the conclusions of Figures \ref{MVE} to \ref{orientation}. Therefore, in the presence of prominent marine wind, it is of vital importance to adapt to or even proactively utilize the wind for energy efficient trajectory design.

\begin{figure}[h]
		\centering
		\includegraphics[width=0.65\linewidth,  trim=0 0 0 0,clip]{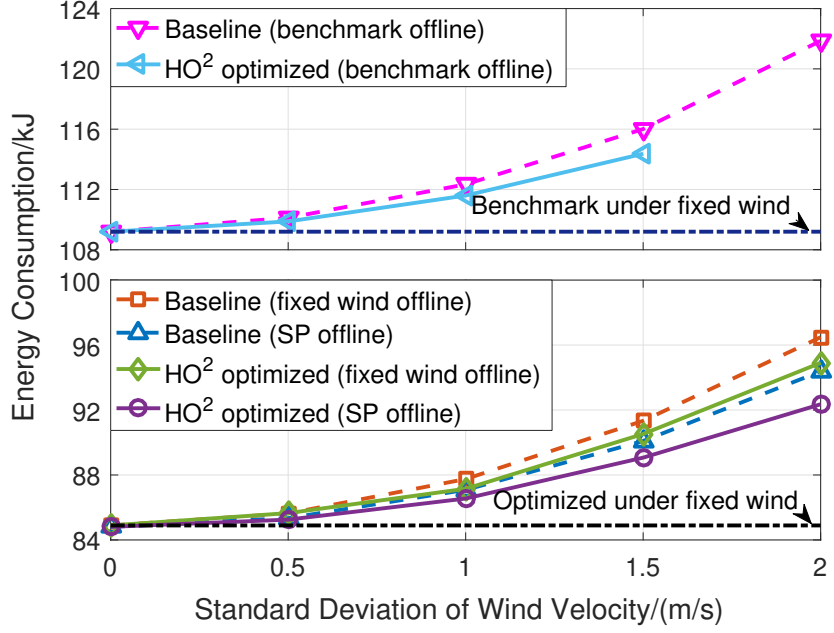}
		\caption{Energy consumption for 8-shape under different standard deviation of wind velocity.\vspace{-1ex}}
		\label{Wind speed variance}
\end{figure}

\paragraph{Wind Variance}
Consider stochastic wind velocity with mean 10 m/s (from south to north) and different standard deviation $\sigma_f$. For illustration, consider 8-shape trajectory with $Q=6$ Gbits and $Q_0=400$ Mbits. The results are plotted in Fig. \ref{Wind speed variance}.

First, for the special case under fixed wind, the energy consumption of our optimized 8-shape trajectory and that of the benchmark trajectory (exact 8-shape with optimized $V$, $r$ and $T_0$) are shown by dash-dotted lines in the lower and upper graphs of Fig. \ref{Wind speed variance}, respectively, whereby the former (around 85 kJ) is significantly lower than the latter (around 109 kJ). This confirms the effectiveness of our offline optimization that minimizes the overall energy consumption.

Second, for the random wind case with different variance, we compare our proposed HO$^2$ design with the baseline scheme which completely follows the offline path without any online adaptation. Since the baseline scheme may violate the airspeed and acceleration constraints, it may not be feasible in online operation.\footnote{In fact, for the example setup here, the baseline scheme almost surely violates one or more constraints along the whole path.}
Nevertheless, for comparison with our proposed HO$^2$ scheme, we still record its theoretical energy consumption regardless of the constraint violation.
From Fig. \ref{Wind speed variance}, it can be seen that the energy consumption increases with wind variance, since a larger wind variance would enlarge the gap between the airspeed online and the optimized airspeed offline.
On the other hand, it can be observed that our proposed HO$^2$ scheme consumes less energy than the baseline while ensuring feasible online operation, and the gap becomes obvious as the wind increases. The underlying reason is that the HO$^2$ scheme is able to refine the trajectory and airspeed in each time slot in adaptation to real-time wind velocity.


Finally, for both the proposed HO$^2$ scheme and the baseline scheme, we consider three types of offline trajectories, i.e., 1) benchmark (exact 8-shape with optimized $V$, $r$ and $T_0$ under fixed wind), 2) optimized under fixed wind (offline optimization based on Section \ref{fixed wind}) and 3) optimized using SP (offline optimization based on Section \ref{offlinedesign}).
Several interesting observations are made as follows. 
First, compared with the benchmark trajectory, the optimized offline trajectory based on fixed wind or SP has much lower energy consumption. Second, the trajectory based on SP offline performs the best, which optimizes all the constraints considering the wind statistics. 
Third, it is worth mentioning that even with online adaptation, the benchmark trajectory may not be feasible when the wind velocity changes greatly (e.g., when $\sigma_f=2.0$ m/s). Therefore, our optimized trajectory can better adapt to the wind variations in practice.

\subsection{Multi-buoy Case}
\subsubsection{Long Buoy Distance}
\begin{figure}[h]
		\centering
		{
			\label{MTrajectory} 
			\includegraphics[height=0.36\linewidth,width=0.46\linewidth]{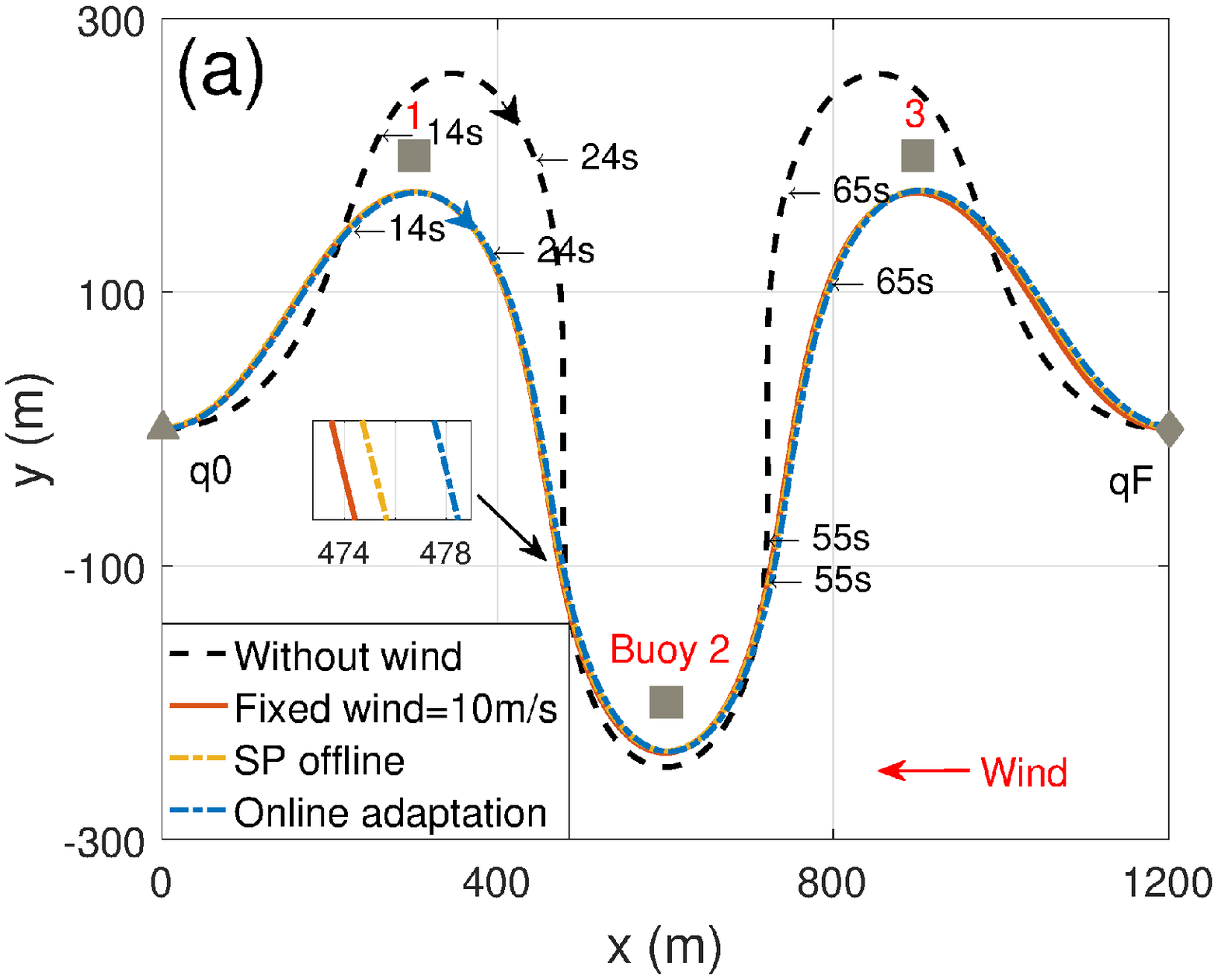}}
		{
			\label{CSMA} 
			\includegraphics[height=0.36\linewidth,width=0.45\linewidth]{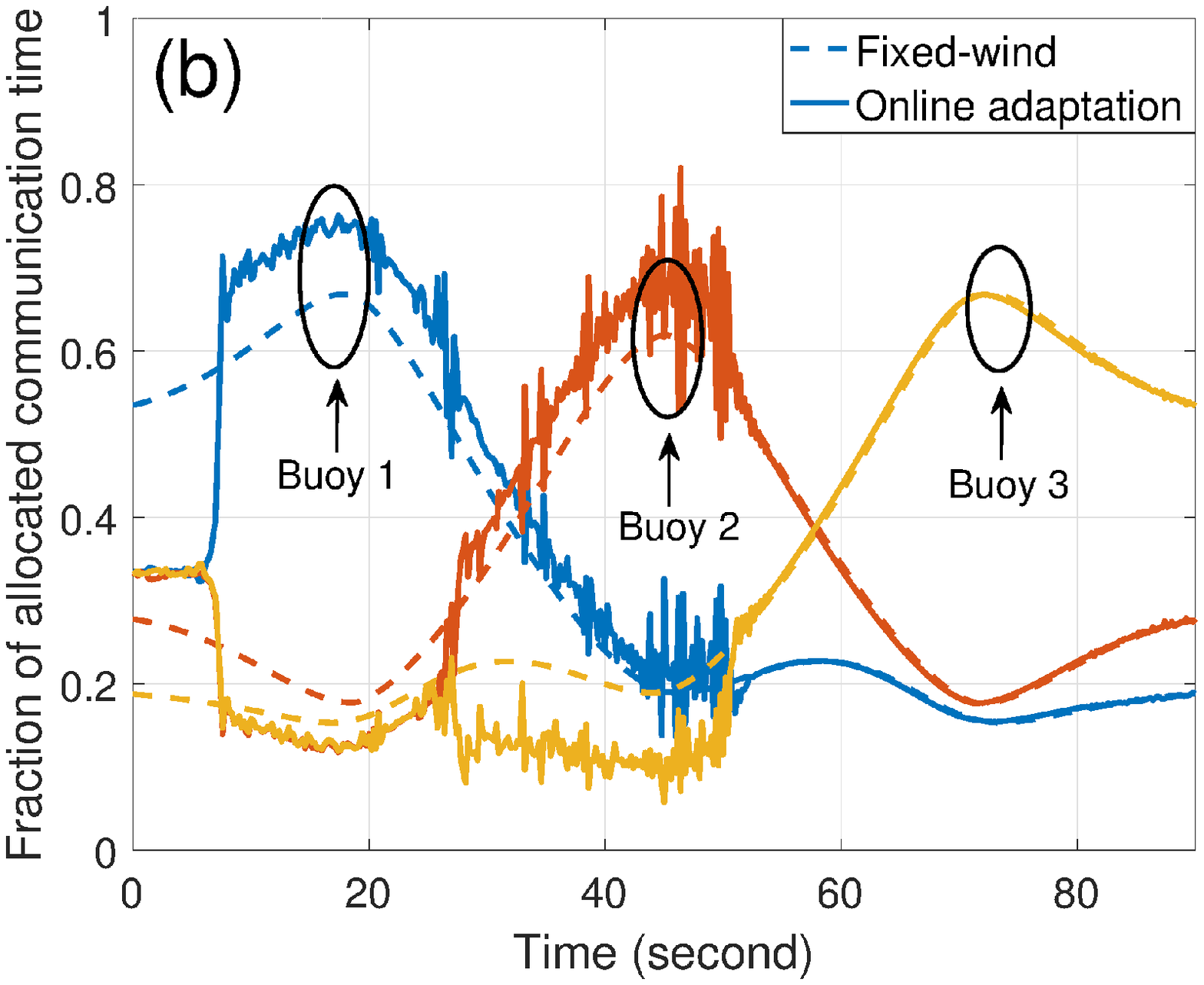}}
		\caption{(a) Optimized UAV trajectory with/without wind. (b) Communication time allocation, with wind (similar for the case without wind).\vspace{-1ex}}
		\label{Multi}
\end{figure}

In this case, the buoys are distributed in a large area and are far from each other.
If the data volume $Q$ to be collected from each buoy is large, we could devise a proper visiting order of the buoys, and then apply our cyclical trajectory design for each one of them; if the data volume $Q$ is small/moderate, we could jointly optimize the trajectory and communication to reduce the UAV's energy consumption, subject to wind effect.
For illustration, consider $\mathbf{q}_{0}={[0,0]^T}$ m, $\mathbf{q}_{F}={[1200,0]^T}$ m, $V_w=10$ m/s (headwind, from east to west), and three buoy locations as shown in Fig. \ref{Multi}(a).
Assume that each buoy has the same required $Q=200$ Mbits and the UAV needs to complete the task in time $T=90$ s.

We first discuss the optimized UAV trajectory and communication time allocation for the fixed-wind situation.
It can be seen from Fig. \ref{Multi}(b) that more time is allocated to a buoy when the UAV flies closer to it, which conforms to the cyclical TDMA principle to exploit the good channel associated with short UAV-buoy distance.
Then, under the headwind condition, by properly optimizing the UAV's trajectory and airspeed, the UAV can have more time to communicate with each buoy at a shorter distance, which helps to shorten the flight trajectory and thus reduce energy consumption by 10.4$\%$ (e.g., 10.49 kJ without wind and 9.40 kJ with wind).
This thus validates the effectiveness of our joint trajectory and communication design in minimizing the UAV's energy consumption subject to wind effect. 
	
Next, we study the effectiveness of the proposed HO$^2$ design for random wind. Assuming standard deviation $\sigma_f$ of wind equal to $1.0$ m/s, the offline optimized trajectory based on SP via SAA is shown by the yellow line while the online fine-tuned trajectory is shown by the blue line in Fig. \ref{Multi}(a). The energy consumption after HO$^2$ optimization is about 9.46$\sim$9.52 kJ, which is still lower than that of the case without wind. Therefore, the online adaptation further refines the UAV airspeed and communication time for feasible online operation as shown in Fig. \ref{Multi}(b), which can also effectively utilize the wind for reducing energy consumption. These results are consistent with those obtained under single-buoy setup.

\subsubsection{Short Buoy Distance}
\begin{figure}[h]
		\centering
		\includegraphics[height=0.61\linewidth,width=0.74\linewidth]{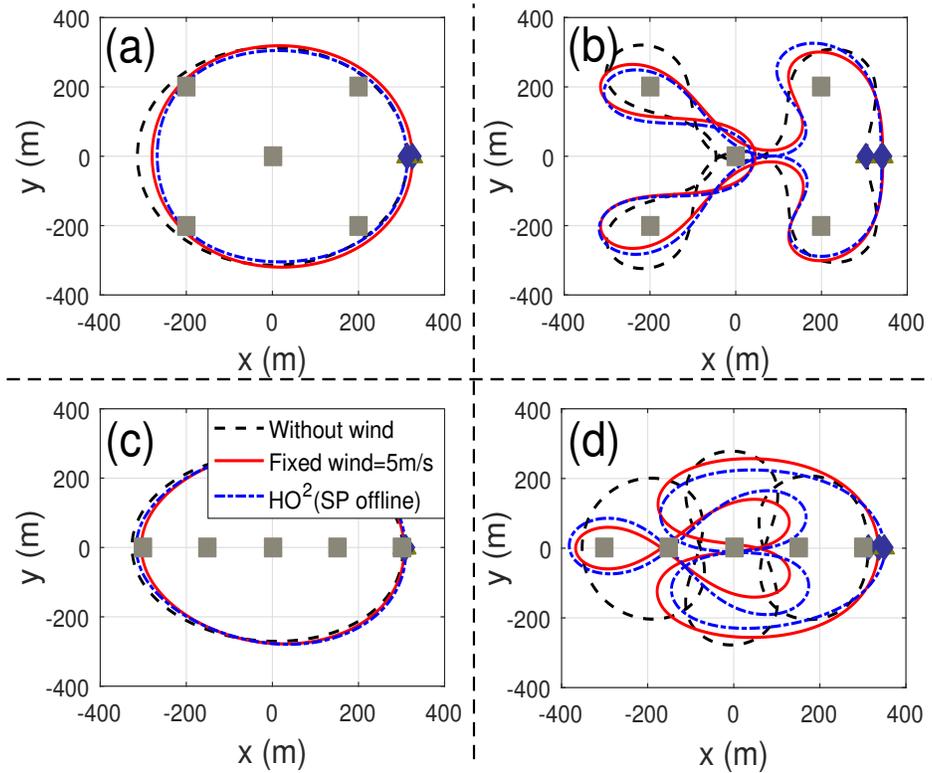}
		\caption{The optimized UAV trajectory for the topology with spreaded buoys with (a) $Q_0=100$ Mbits or (b) $Q_0=200$ Mbits, and for the line topology with (c) $Q_0=100$ Mbits or (d) $Q_0=200$ Mbits.\vspace{-2ex}}
		\label{MB}
\end{figure}

In this case, the buoys are clustered in a small region.
We can then design cyclical trajectories to fit the buoys' topology and also cater for the wind effect.
For illustration, consider two topologies each with five buoy locations shown in Fig. \ref{MB}(a) and (c), respectively.
Assume that the wind blows from south to north. For the case with large data volume to be collected, we consider two kinds of data partitions with $Q_0=100$ Mbits and $Q_0=200$ Mbits in each lap, respectively. 
The optimized UAV trajectories for different buoy topology and $Q_0$ with/without wind are shown in Fig. \ref{MB}. 
	
For the topology in Fig. \ref{MB}(a) and (b) with spreaded buoys,
as $Q_0$ increases, it is observed that the optimized trajectory gets closer to each of the buoys in order to collect data at a higher rate.
Similar results are observed for the line topology in Fig. \ref{MB}(c) and (d).
In addition, by comparing Fig. \ref{MB}(a) and (c) under the same $Q_0$, it is observed that the optimized trajectory becomes flat in Fig. \ref{MB}(c), which tends to fit the buoys' topology and get closer to the buoys.
The above results further validate our proposed joint trajectory and communication design in adapting to the buoys' topology under different data volume requirement. Meanwhile, similar to the single-buoy case, the optimized trajectory can proactively exploit the wind and achieve lower energy consumption.
For instance, consider the topology with spreaded buoys with $Q_0=200$ Mbits in Fig. \ref{MB}(b).
We compare our optimized trajectory with the benchmark trajectory (not shown in the figure for brevity) which is exact circular with optimized $V$, $r$, $T$ and equal time allocation for each buoy. First, for the case without wind, it is observed that the energy consumption of our optimized trajectory is much lower than the benchmark (e.g., 19.50 kJ for optimized scheme and 26.51 kJ for benchmark). Second, for the fixed-wind case, the optimized trajectory can even utilize the wind to further reduce the energy consumption, while the benchmark scheme experiences increase in the consumed energy (e.g., 18.36 kJ for optimized scheme and 27.25 kJ for benchmark).
Finally, for the case with random wind, the proposed HO$^2$ design can still adapt to the wind variations effectively in these multi-buoy setups with low energy consumption (e.g., 18.45 kJ$\sim$18.74 kJ with $\sigma_f=1.0$ m/s). 
	
\section{Conclusions}\label{conclusions}
This paper investigates a maritime data collection system with a fixed-wing UAV dispatched as a mobile data collector, and aims to minimize its energy consumption by joint trajectory and communications optimization, subject to marine wind effect.
This problem is non-convex and difficult to solve, especially when the target data volume is large.
We propose a new cyclical trajectory design framework that can handle arbitrary data volume efficiently subject to wind effect, with reduced trajectory/computational complexity.
Furthermore, we propose the HO$^2$ optimization which adapts to wind variations by leveraging both the statistical and real-time wind velocity.
Numerical results show that our proposed cyclical scheme can proactively utilize the wind and fit the buoys' topology, which achieves significant energy savings compared with benchmark schemes. Moreover, the HO$^2$ design can effectively adapt to wind variations with feasible and robust online operation, as well as low energy consumption.
Finally, the proposed HO$^2$ optimization framework is general and can potentially be applied to other scenarios with stochastic environmental factors.

\bibliography{IEEEabrv,BibDIRP}
	
\newpage
\ifCLASSOPTIONcaptionsoff
\newpage
\fi
	
\end{document}